\documentclass[verbose,letterpaper,margin=1in]{article}
\usepackage[final]{neurips_2025}
\usepackage[utf8]{inputenc}
\usepackage[T1]{fontenc}
\usepackage{url}
\usepackage{wallpaper}
\usepackage{booktabs}
\usepackage{array}
\usepackage{float}
\usepackage{caption}
\usepackage{tikz}
\usetikzlibrary{shapes,arrows,positioning}
\usetikzlibrary{shapes.geometric, backgrounds, arrows.meta, shapes.symbols, fit}
\usepackage{fontawesome5}
\usepackage{qrcode}
\usetikzlibrary{shapes.callouts}
\usepackage{booktabs}
\usetikzlibrary{calc}
\usepackage{amsfonts}
\usepackage{nicefrac}
\usepackage{microtype}
\usepackage{xcolor}
\usepackage{graphicx}
\usepackage{longtable}
\usepackage{amsmath}
\usepackage{amsmath}
\usepackage{amssymb}
\usepackage{fancybox} 
\usepackage{enumitem}
\usepackage{float}
\usepackage{caption}
\usepackage{xcolor,colortbl}
\usepackage{pifont} 
\newcommand{\cmark}{\textcolor{green!60!black}{\ding{51}}}
\newcommand{\xmark}{\textcolor{red!70!black}{\ding{55}}}
\usepackage{hyperref} 

\title{The Trust Fabric: Decentralized Interoperability and Economic Coordination for the Agentic Web}

\author{%
  Sree Bhargavi Balija \\
  UC San Diego \\
  \texttt{sbalija@ucsd.edu} \\
  \And
  Rekha Singal \\
  TCS \\
  \texttt{rekha.singhal@tcs.com} \\
  \And
  Ramesh Raskar \\
  MIT Media Lab \\
  \texttt{raskar@mit.edu} \\
    \And
    Erfan Darzi
 \\
  Harvard University \\
  \texttt{darzi@mit.edu} \\
    \And
   Raghu Bala \\
  Synergetics \\
  \texttt{} \\
   \And  
   Thomas Hardjono \\
  MIT Connection Science \\
  \texttt{} \\
   \And  
   Ken Huang \\
   Agentic AI Security \\
\texttt{} \\
}
\makeatletter
\def\@trackname{} 
\makeatother
\begin{document}
\maketitle

\begin{abstract}
The fragmentation of AI agent ecosystems has created urgent demands for interoperability, trust, and economic coordination that current protocols (MCP \cite{hou2025,desai2025}, A2A \cite{habler2025}, ACP \cite{liu2025}, and Cisco's AGP \cite{edwards2025}) cannot address at scale. We present the Nanda Unified Architecture, a decentralized framework built around three core innovations: fast DID-based agent discovery through distributed registries enables efficient lookup across decentralized networks, while semantic agent cards with verifiable credentials and composability profiles provide rich, machine-readable descriptions of capabilities. At the heart of the system, a dynamic trust layer integrates behavioral attestations with policy compliance mechanisms to create verifiable reputation signals. The architecture introduces X42/H42 micropayments for economic coordination and MAESTRO, a comprehensive security framework incorporating Synergetics' patented AgentTalk protocol (US 12,244,584 B1) and secure containerization. Real-world implementations demonstrate 99.9\% compliance in healthcare applications and significant monthly transaction volumes while maintaining strong privacy guarantees. Our federated registry system enables efficient agent discovery while supporting high-performance autonomous systems. By unifying MIT's trust research with production systems from Cisco's Agency Framework and Synergetics' commercial deployments, we demonstrate how cryptographic proofs and policy-as-code transform agents into trust-anchored participants in a decentralized economy \cite{lakshmanan2025,sha2025}. The result enables a globally interoperable Internet of Agents where trust becomes the native currency of collaboration across both enterprise and Web3 ecosystems.
\end{abstract}

\section{Introduction}
The AI ecosystem is undergoing a rapid transformation, with autonomous agents emerging as the fundamental units of intelligence across both consumer applications and enterprise-grade workflows. Just as containerization and Kubernetes revolutionized cloud-native computing by standardizing deployment and orchestration, a parallel shift is now unfolding in the agentic landscape. Agent orchestration is becoming the next abstraction layer—enabling scalable, intelligent, and adaptive multi-agent systems that can reason, collaborate, and act autonomously.

\begin{figure}[htbp]
  \centering
  \setlength{\fboxsep}{4pt}   
  \setlength{\fboxrule}{0.8pt} 
  \fbox{\includegraphics[height=0.3\textheight]{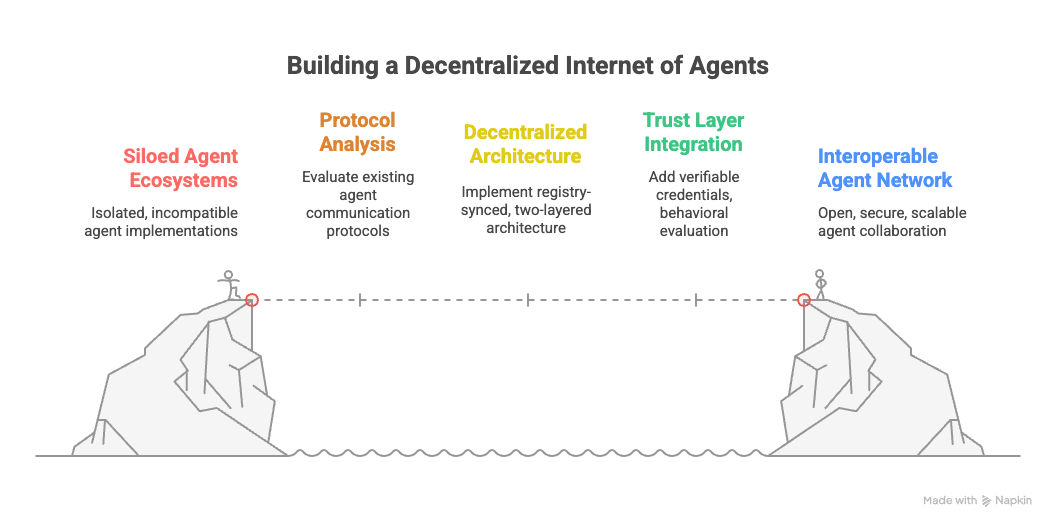}}
  \caption{Building decentralized internet of agents}
  \label{fig:architecture}
\end{figure}

To ground this shift in a systems-level context, Figure~\ref{fig:architecture} presents a high-level overview of the Nanda Unified Architecture. This architecture lays the foundation for a globally interoperable agent economy by introducing layered abstractions for discovery, composition, evaluation, incentivization, and deployment. Each layer addresses specific bottlenecks in current agent-based design patterns and is intended to work in synergy with decentralized registries and trust mechanisms.

Despite this growing enthusiasm, the current agent landscape remains fragmented. Protocols, toolchains, registries, and incentive structures are often siloed, hindering agent-to-agent collaboration and monetizable cooperation. Without a shared infrastructure for identity, discovery, trust, and reputation, the ecosystem risks echoing the pitfalls of early closed and incompatible AI silos.

Recent initiatives such as the Model Context Protocol (MCP) \cite{hou2025,desai2025,lakshmanan2025}, Agent-to-Agent Protocol (A2A) \cite{habler2025}, and Agent Connect Protocol (ACP) \cite{liu2025} have introduced promising abstractions for communication. However, these protocols primarily target execution orchestration and fail to adequately address the deeper infrastructural needs of agent discovery, semantic identity, and dynamic trust management at scale.

This paper argues that enabling seamless collaboration between billions of autonomous agents requires foundational primitives that go beyond basic messaging. In particular, we advocate for a robust trust layer that quantifies, validates, and maintains agent reputation and behavioral integrity.

To address this, we build upon the contributions of the Nanda research collective at MIT, in collaboration with Synergetics, Cisco, Flower, Dell, HCL, TCS, and other academic partners. We propose a two-layered registry architecture tailored for the Agentic Web:

\begin{itemize}
    \item \textbf{Layer 1:} A lightweight, fast-resolving registry that maps agent names or decentralized identifiers (DIDs) to metadata URLs, supporting rapid resolution and lookup.
    \item \textbf{Layer 2:} A semantic agent card (or "agent fact") layer that extends A2A's foundational ideas to include verifiable credentials, composability profiles, adaptive routing metadata, and service history.
\end{itemize}
\begin{figure}[htbp]
  \centering
  \setlength{\fboxsep}{4pt}   
  \setlength{\fboxrule}{0.8pt} 

  \begin{minipage}[b]{0.45\textwidth}
    \centering
    \fbox{\includegraphics[width=\textwidth,height=1.05\textwidth]{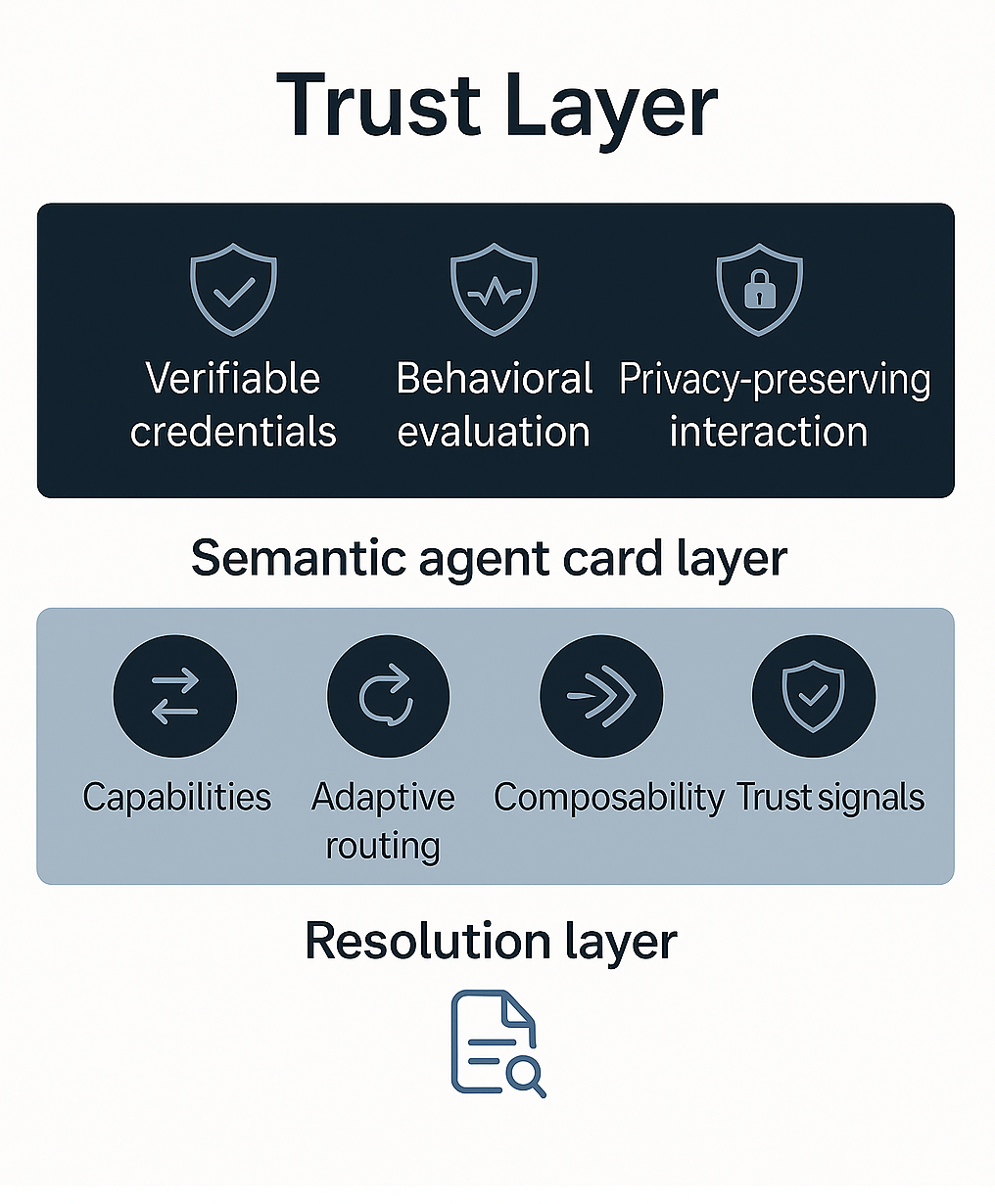}}
    \caption*{(a) Trust layer-enabled agent stack}
  \end{minipage}
  \hspace{0.05\textwidth}
  \begin{minipage}[b]{0.45\textwidth}
    \centering
    \fbox{\includegraphics[width=\textwidth,height=1.05\textwidth]{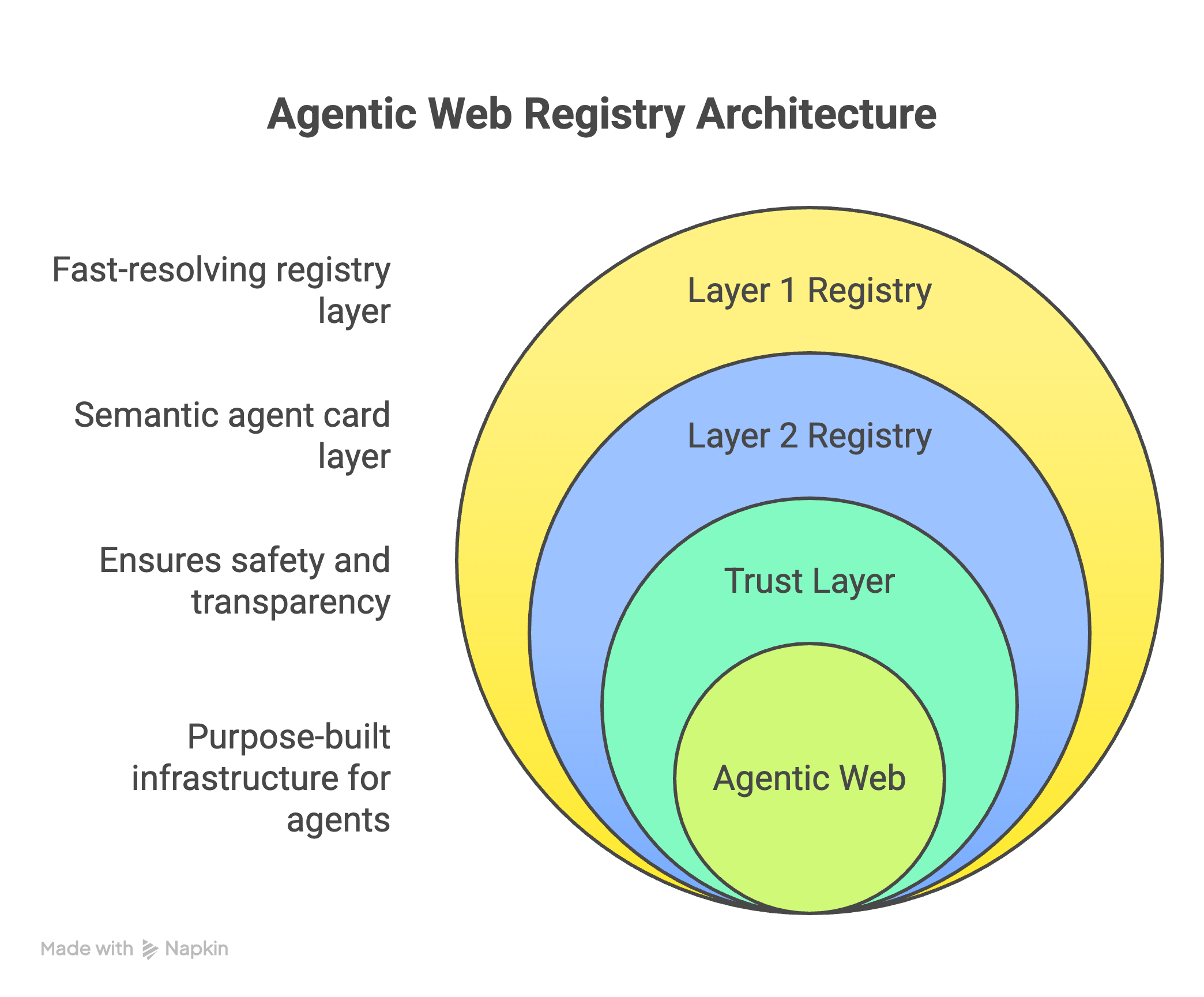}}
    \caption*{(b) Registry-synced agent discovery}
  \end{minipage}

  \caption{Architectural illustrations of agent stack and discovery layer}
  \label{fig:trust-discovery}
\end{figure}

These two registry layers are unified via a trust layer that employs distributed and federated trust \cite{a1,a2,a3,a4,a5,a6} authorities, behavioral evaluation engines, and credential-based attestations. This design ensures that agents operate within safe, transparent, and verifiable boundaries.

Inspired by the shift from dial-up to broadband, we argue that the Agentic Web necessitates a purpose-built infrastructure where semantic discoverability, programmable incentives, and behavioral safety become first-class primitives. Furthermore, perspectives from Mayfield Ventures, Acorn Labs and Vigil \cite{edwards2025, sha2025} emphasize the need for trustworthy, auditable, and test-driven agent deployments, particularly in mission-critical domains such as healthcare, finance, and defense. As Vin Sharma from Vigil observes, the industry must shift from demo-ready chatbots to production-grade autonomous systems—systems that not only follow strict policy constraints and pass rigorous behavioral audits, but also remain safe and predictable even when operating in adversarial or unexpected conditions. Instead of crashing, malfunctioning, or producing unsafe outputs, such agents should exhibit controlled, measurable responses—maintaining baseline functionality while signaling failure modes transparently. This form of robustness is essential for real-world deployment.

Through this paper, we aim to:
\begin{enumerate}
    \item Provide a unified perspective on registry and trust-layer requirements for agentic systems across consumer and enterprise use cases.
    \item Evaluate the ``upgrade vs. switch'' paradigms in adapting today's web infrastructure for decentralized agent interactions.
    \item Propose design principles and open questions for building a resilient, incentivized, and composable Internet of Agents.
\end{enumerate}

Ultimately, we envision a future where agents are not only interoperable but also economically and behaviorally aligned operating across a globally distributed mesh of registries, protocols, and verifiable trust layers.
 
\begin{figure}[htbp]
  \centering
  \includegraphics[width=1.1\textwidth, height=0.4\textheight]{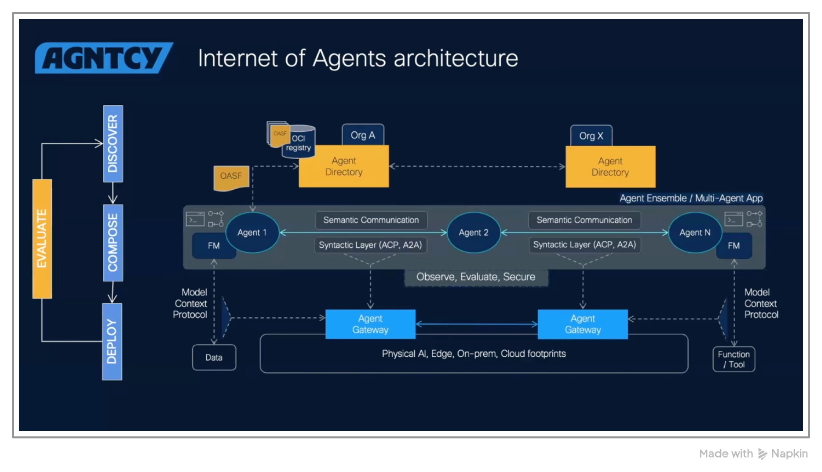}
  \caption{Internet of Agents: A Blueprint for Agent Collaboration and Secure Communication}
  \label{fig:architecture1}
\end{figure}

\section{Related Work and Protocol Landscape}

The evolution of agent interoperability protocols reflects an accelerating recognition that agent ecosystems must be both decentralized and composable. Rather than listing protocol features in isolation, this section synthesizes existing efforts into a coherent landscape of architectural primitives, drawing attention to their conceptual overlaps, implementation gaps, and collective momentum toward a more robust agentic future.

Early agent interoperability protocols such as the \textbf{Model Context Protocol (MCP)} were introduced to enable structured context exchange between AI agents and external tools or data sources \cite{ehtesham2025, kumar2025}. MCP gained early adoption due to its lightweight design and practical utility in model pipelines, particularly for fine-tuning context-aware behavior. However, its reliance on static context schemas and manual endpoint registration limits its scalability in dynamic, peer-to-peer agent ecosystems and hinders real-time orchestration in decentralized settings \cite{edwards2025, sha2025}.

In response, the \textbf{Agent-to-Agent Protocol (A2A)} introduced self-declared agent cards and human-readable descriptions to facilitate basic agent discoverability and interaction \cite{habler2025}. While a meaningful step forward, A2A struggles with lack of runtime metadata and composability hooks, constraining its use in high-frequency, multi-agent coordination scenarios \cite{liu2025}. Commercial implementations like \textbf{AgentTalk (Synergetics)} extend A2A with decentralized identity (DIDs) and micropayments, though they introduce vendor-specific considerations.

Building upon these foundations, Cisco's \textbf{Agency protocols} notably the \textbf{Agent Connect Protocol (ACP)} and \textbf{Agent Gateway Protocol (AGP)} propose a more dynamic and interoperable model \cite{tran2025, lyu2025}. These protocols explicitly target decentralized agent ecosystems, enabling agent-to-agent communication, group coordination, and vendor-agnostic composability. By treating agent interactions as first-class entities rather than repurposed tool integrations, ACP and AGP reflect a shift toward more expressive communication layers.

Complementing these communication protocols are efforts to formalize agent registration and discovery through federated registries. Cisco's \textbf{Open Agentic Schema Framework (OASF)} and decentralized \textbf{Agent Directories} leverage OCI-like structures to define interoperable schemas and registries for agent metadata \cite{hou2025}. These initiatives align with proposals from the Nanda registry architecture, which advocates for a globally distributed "quilt" of registries—hybrid in nature, spanning both enterprise institutions and Web3-native communities \cite{wiggers2025a, jackson2025}. \textbf{Synergetics} operates a production-ready AgentRegistry implementing NANDA's DID-based schema, demonstrating how academic research (MIT Media Lab) can bridge to enterprise adoption through decentralized agent discovery and verification services.

Underlying many of these proposals is the growing influence of \textbf{Decentralized Identity (DID)} standards and Web3 trust primitives. These frameworks provide essential infrastructure for verifiable agent identity, policy enforcement, and privacy-preserving interactions \cite{masson2025, janakiram2024}, aligning with the dual imperative of serving both regulated enterprise applications and open consumer networks.

Taken together, these protocols and frameworks signal a shared trajectory: toward agent-native infrastructure that is decentralized, trust-rich, and economically aligned. Our proposed architecture builds directly upon this foundation by integrating trust as a first-class design layer—enabling verifiable credentials, behavioral attestations, and secure, privacy-respecting agent orchestration across domains.
\begin{figure}[htbp]
  \centering
  \fbox{\includegraphics[height=0.3\textheight]{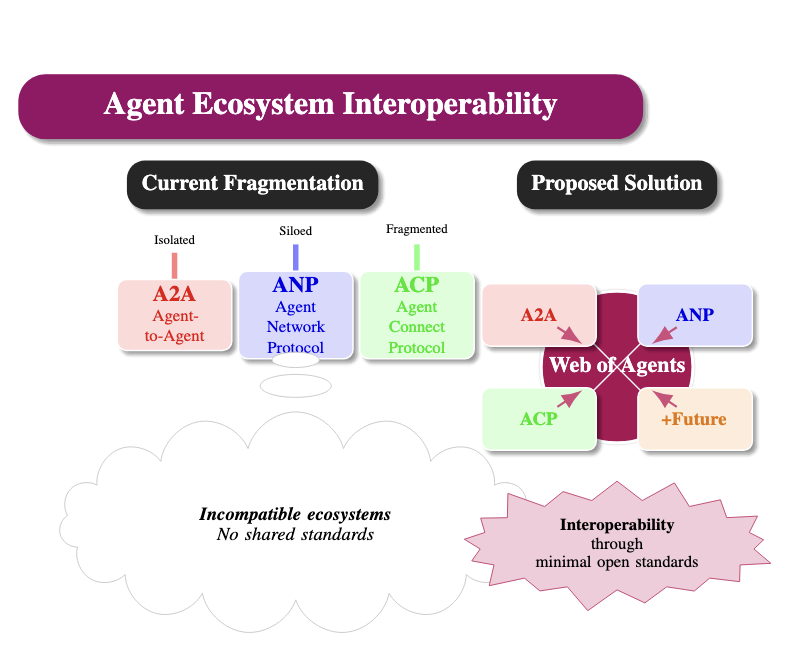}}
\end{figure}
\subsection{Protocol Composition Algebra}
The NANDA architecture enables \textit{protocol gene splicing} through operator composition:

\begin{equation*}
\underbrace{\text{DID}}_{\substack{\text{Decentralized}\\\text{Identity}}} \oplus 
\underbrace{\text{VC}}_{\substack{\text{Verifiable}\\\text{Credentials}}} \otimes 
\underbrace{\text{X42}}_{\substack{\text{Atomic}\\\text{Payments}}} = 
\text{Agent Gene}
\end{equation*}

where $\oplus$ denotes identity-binding and $\otimes$ represents incentive-aligned composition. Synergetics' AgentTalk implements this as:
\begin{equation*}
\text{Agent Gene} \bowtie \text{Policy} \rightarrow \text{Trusted Agent}
\end{equation*}
\vspace{1em}

\renewcommand{\arraystretch}{1.2} 
\begin{table}[htbp]
\centering
\caption{Comparison of Protocols in Agent Interoperability Landscape}
\begin{tabular}{p{2.5cm} p{3.2cm} p{5.5cm} p{4.5cm}}
\toprule
\textbf{Protocol Name} & \textbf{Origin} & \textbf{Limitations} & \textbf{Use Cases} \\
\midrule
AgentTalk & Synergetics (Patented: US 12,244,584 B1) & Vendor-specific implementation; requires DID adoption; enterprise-focused pricing structure & Enterprise A2A workflows with micropayments; merchant-agent integration; marketplace transactions \\
\addlinespace
MCP & Model-Tool Communication Protocol & Lacks dynamic agent-to-agent orchestration; manual tool lookup; not suited for scalable peer-to-peer agent collaboration & Tool orchestration in data pipelines; agent execution within ML workflows \\
\addlinespace
A2A & Agent-to-Agent Protocol & Lacks dynamic behavioral metadata; not robust for large-scale multi-agent scenarios & Initial prototypes for agent discoverability; small-scale interoperability \\
\addlinespace
ACP & Cisco's Agency & Execution-level focus; may require further extension for semantic and trust-layer integration & Mission-critical workflows; cross-enterprise orchestration \\
\addlinespace
AGP & Cisco's Agency & Early-stage protocol; may require integration with dynamic discovery and adaptive trust layers & Secure enterprise-grade agent group communication \\
\addlinespace
OASF & Cisco's Agency & Focuses on metadata and schema; requires complementary trust layer for holistic safety guarantees & Extending agent descriptions for security audits and trust alignment \\
\bottomrule
\end{tabular}
\label{tab:protocols}
\end{table}
\subsection{Agent-to-Merchant Communication Framework}
\label{sec:agent_merchant}

The consumerization of agent-to-merchant communication enables seamless interactions through standardized protocols as shown in below figure.

\begin{figure}[htbp]
\centering
\begin{tikzpicture}[
    node distance=1cm,
    agent/.style={rectangle, rounded corners, draw=blue!50, fill=blue!10, minimum width=3cm},
    merchant/.style={rectangle, rounded corners, draw=green!50, fill=green!10, minimum width=3cm},
    protocol/.style={ellipse, draw=orange, dashed, fill=orange!5},
    arrow/.style={->, >=stealth, thick}
]

\node (customer) [agent] {Customer Agent {\scriptsize \textcolor{blue}{(Agent)}}};
\node (api) [below=of customer, merchant] {Merchant Hub {\scriptsize \textcolor{green}{(API + Wallet)}}};
\node (wizard) [left=of api, merchant, xshift=-2cm] {Agent Wizard};
\node (magent) [below=of api, merchant] {Merchant Agent {\scriptsize \textcolor{green}{(Agent)}}};
\node (a2a) [protocol, right=of api, xshift=2cm] {AgentTalk (A2A) {\scriptsize \textcolor{orange}{(Secure)}}};

\draw [arrow] (customer) -- node[right] {API Calls} (api);
\draw [arrow] (wizard) -- node[above] {Onboarding} (api);
\draw [arrow, bend left=20] (customer) to node[near start, right] {Protocol} (a2a);
\draw [arrow, bend right=20] (a2a) to node[near end, right] {A2A} (magent);
\draw [arrow] (api) -- (magent);

\node [below right=0.5cm and -3cm of magent, text width=4cm] {
    \footnotesize
    \textcolor{blue}{$\blacksquare$} Customer-side \\
    \textcolor{green}{$\blacksquare$} Merchant-side \\
    $\blacktriangleleft$ Proprietary communication
};
\end{tikzpicture}
\caption{Agent to Merchant Communication}
\label{fig:agent_communication} 
\end{figure}
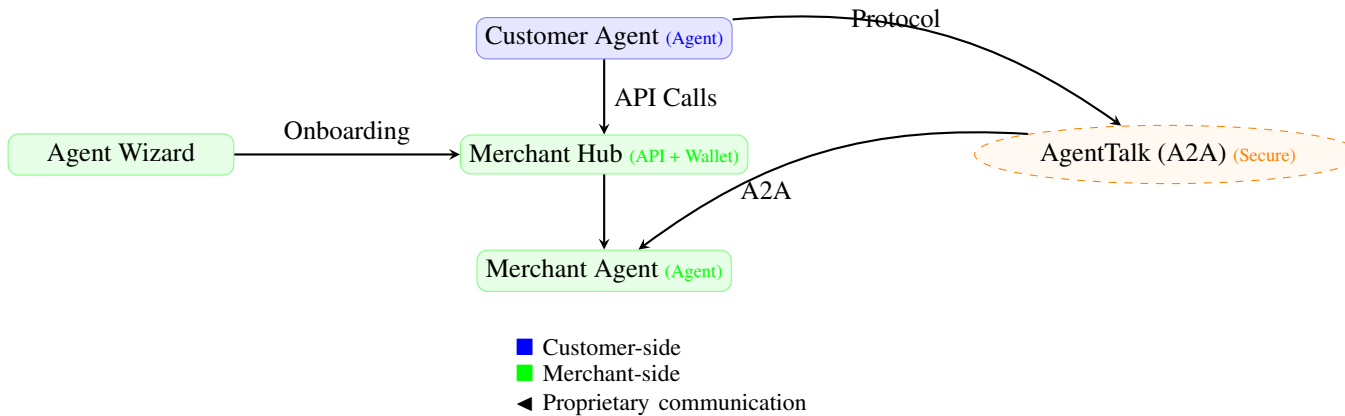

\subsection{Architecture Components}
The Agent to Merchant Communication framework consists of three core components:

\begin{itemize}
    \item \textbf{Merchant MCP Server}: Hosted infrastructure with existing APIs and wallet functionality
    \item \textbf{Agent Wizard}: Onboarding platform for merchant agents
    \item \textbf{AgentTalk Protocol (A2A)}: Standardized communication protocol
\end{itemize}

\subsection{Communication Flow}
The interaction sequence follows:

\begin{enumerate}
    \item Customer agents initiate via API calls to merchant MCP servers
    \item Merchant agents are onboarded through Agent Wizard
    \item Both parties leverage AgentTalk (A2A) for secure communication
\end{enumerate}

\subsection{Advantages}
Key benefits include:

\begin{itemize}
    \item Standardized interface preserving existing infrastructure
    \item Secure communication channel (as shown in proprietary implementation)
    \item Simplified agent onboarding process
\end{itemize}
\subsection{Mathematical Models and Formal Methods for Trust, Incentivization, and Interoperability}

To ensure secure interoperability, trustworthy agent behavior, and efficient coordination within the Nanda architecture, we adopt formal mathematical models. These models span privacy-preserving computation, registry synchronization efficiency, and graph-theoretic trust propagation. Each subsection below introduces a model, its associated parameters, and the relevance to decentralized agent networks.

\subsubsection{Secure Communication and Data Privacy}

Privacy guarantees for agent-internal computations are formalized using differential privacy. Let $\mathcal{M}$ represent a randomized mechanism (e.g., homomorphic encryption operations or zero-knowledge proof generation) operating on an agent's dataset $D$. For any two neighboring datasets $D$ and $D'$, and for any subset of outputs $S$, the mechanism $\mathcal{M}$ satisfies $(\epsilon, \delta)$-differential privacy if:

\begin{equation}
\Pr[\mathcal{M}(D) \in S] \leq e^{\epsilon} \Pr[\mathcal{M}(D') \in S] + \delta
\end{equation}

Here:
\begin{itemize}
    \item $\epsilon$ controls the privacy loss (lower implies stronger privacy).
    \item $\delta$ bounds the probability that privacy is compromised.
    \item $D$ and $D'$ differ by at most one element, ensuring resilience to small input changes.
\end{itemize}

In the Nanda architecture, this model supports privacy-preserving distributed computation, where sensitive agent data must remain confidential even during collaborative operations.

\subsubsection{Dynamic Registry Resolution Time}

Efficient agent registry resolution is critical for scalable multi-agent coordination. When registries are implemented using balanced tree structures such as Merkle tries or radix trees, the time complexity to resolve an agent identity scales logarithmically with the number of registered agents $N$:

\begin{equation}
\text{Resolution Time} = O(\log N)
\end{equation}

In decentralized settings that use eventually consistent registries, such as those based on Conflict-Free Replicated Data Types (CRDTs) or gossip-based synchronization protocols, the time to reach convergence among all registry replicas also exhibits logarithmic behavior:

\begin{equation}
T_{\text{convergence}} = O(\log N)
\end{equation}

This reflects typical performance in peer-to-peer dissemination networks, where the number of rounds required for complete propagation grows sublinearly with network size.

\subsubsection{Trust Score as a Weighted Graph Centrality}

Trust relationships among agents are modeled as a weighted directed graph $G = (V, E)$, where:
\begin{itemize}
    \item $V$ denotes agents.
    \item $E$ denotes directed edges with trust weights $w_{ij}$ indicating the trust agent $i$ places in agent $j$.
\end{itemize}

A local trust score for agent $i$ can be computed as the average weight of incoming trust edges:

\begin{equation}
\text{TrustScore}_i = \frac{1}{d_i} \sum_{j \in N(i)} w_{ij}
\end{equation}

where $N(i)$ is the set of agents that $i$ trusts and $d_i$ is the degree (number of outgoing trust connections) of agent $i$.

To model trust propagation through the entire agent network, we extend this model using a variant of the PageRank algorithm:

\begin{equation}
\mathbf{T} = \alpha \mathbf{W} \mathbf{T} + (1 - \alpha) \mathbf{e}
\end{equation}

where:
\begin{itemize}
    \item $\mathbf{W}$ is a row-stochastic matrix derived from normalized trust weights.
    \item $\alpha \in (0,1)$ is a damping factor that controls the balance between propagated and base trust.
    \item $\mathbf{e}$ is the base trust vector (e.g., uniform or application-specific priors).
\end{itemize}

This recursive formulation captures transitive trust and enhances robustness to manipulation, such as Sybil attacks, by leveraging the topology of the trust graph.

\begin{figure}[htbp]
  \centering
  \setlength\fboxsep{4pt}
  \setlength\fboxrule{0.8pt}
  \fbox{\includegraphics[height=0.3\textheight]{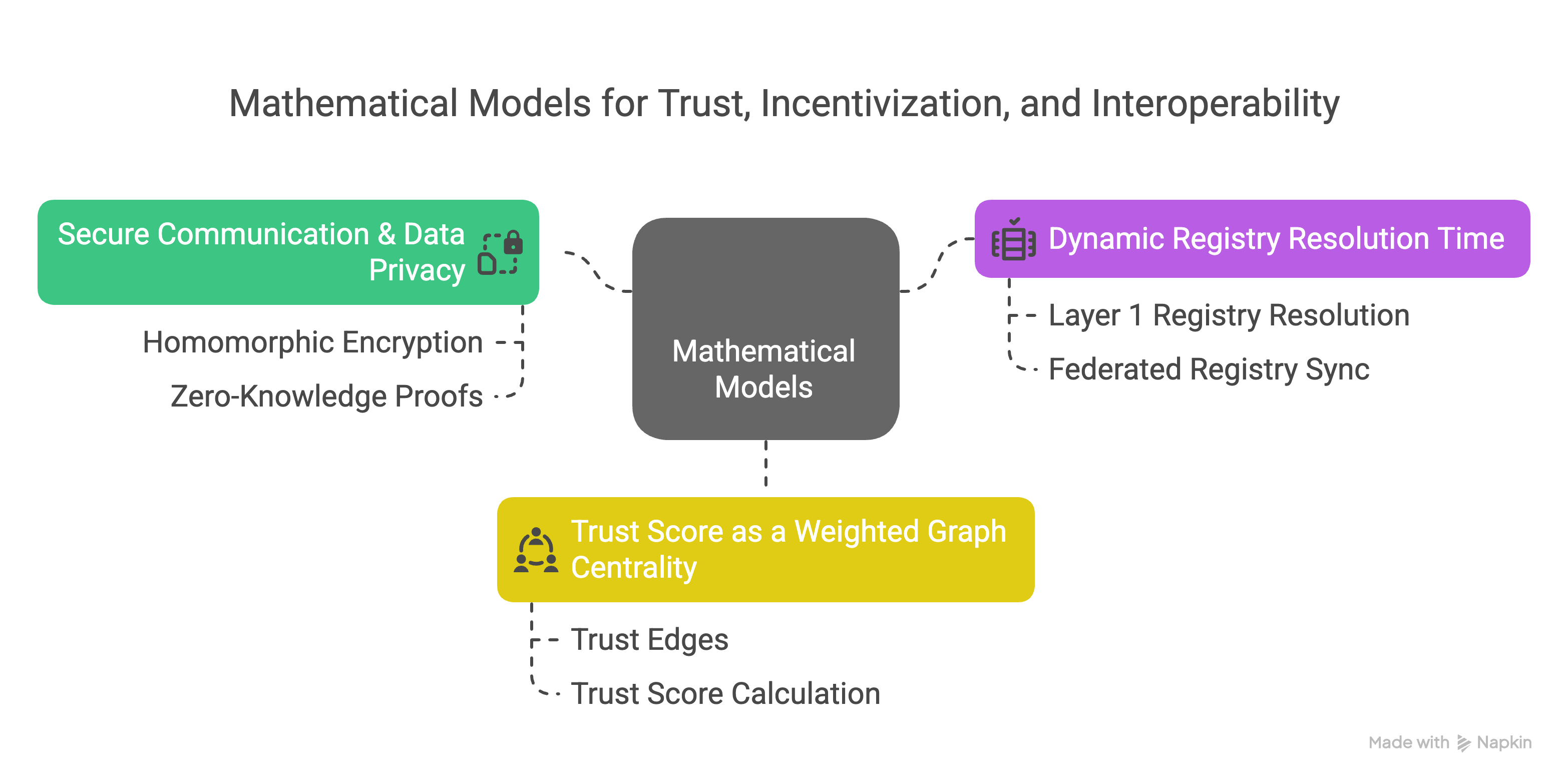}}
  \caption{Illustration of formal models used for privacy, trust, and synchronization in agent networks.}
  \label{fig:architecture8}
\end{figure}

\section{Incentivization: The Role of Microtransactions}

Economic incentivization is a cornerstone of functional decentralization. As demonstrated by Coinbase's X42 protocol, the viability of agent-based ecosystems hinges on native support for real-time microtransactions. These mechanisms enable agents to autonomously compensate one another for services such as computational work, storage provisioning, and API access. For instance, microtransaction enabled communication allows one agent to delegate a task to another and pay instantly upon completion, or to meter and pay for memory and API usage on a per-request basis. Additionally, agents performing machine learning inference tasks can be compensated in real time, encouraging the deployment of high-value, low-latency models within the ecosystem. Unlike traditional payment systems—such as credit cards, which introduce friction through identity verification, settlement delays, and transaction fees, microtransactions via protocols like X42 (transmitted, for example, through HTTP headers) offer lightweight, low-latency economic coordination aligned with the dynamic needs of autonomous agents.
\usetikzlibrary{arrows.meta, shapes.geometric, positioning, backgrounds, fit, shadows}

\begin{figure}[htbp]
  \centering
  \includegraphics[height=0.3\textheight]{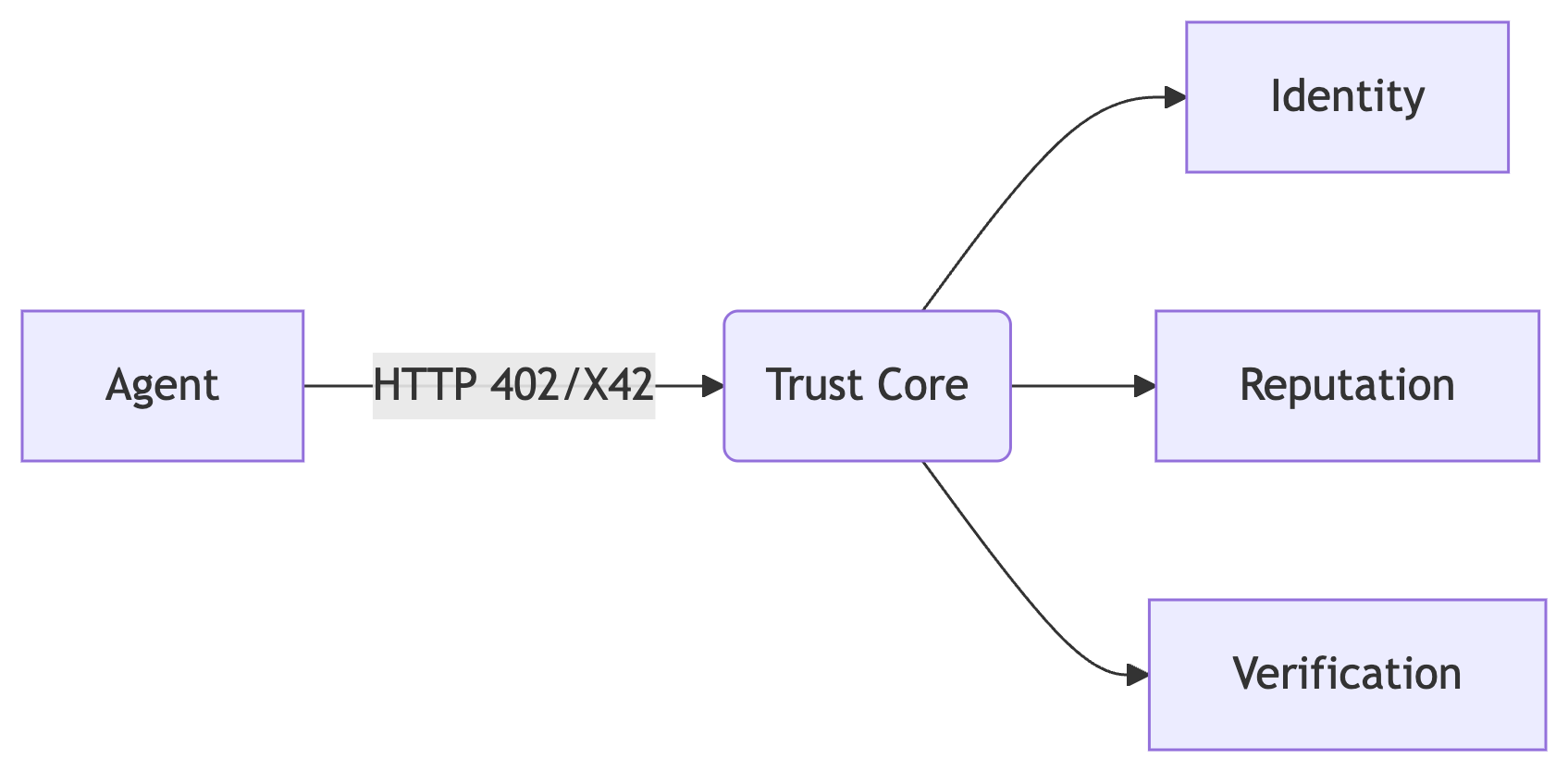}
  \caption{Architecture overview of trust layer-enabled agent stack}
  \label{fig:architecture6}
\end{figure}

\section{Evolving Trust Layers: From HTTP 402 to X42/H42 and Beyond}

The Agentic Web represents a paradigm shift where autonomous agents seamlessly interact and transact across distributed ecosystems. This evolution is underscored by the journey from HTTP 402, a once-forgotten status code reserved for payments, to the modern micropayment protocols X42 and H42 that enable real-time, decentralized agent-to-agent (A2A) transactions. These payment rails act as the foundation for a trust layer, a dynamic construct integrating identity, reputation, and verification modules to ensure reliable, autonomous cooperation. 

As agents independently access data, compute, and intelligence resources, the trust layer expands beyond traditional monetary exchanges to include data/ compute pricing and the emerging notion of knowledge commoditization. In this agentic economy, trust is no longer a passive backdrop but an active participant, enabling agents to self-organize, monetize interactions, and drive discovery without compromising privacy or security. 

The conceptual diagram captures this progression, illustrating how the reimagining of payments and identity layers forms the robust fabric of the Agentic Web, paving the way for packet-switched intelligence and the commoditization of micro-wisdom. 

\captionsetup[table]{skip=10pt}

\begin{table}[H]
\centering
\caption{Trust Evaluation Dimensions for Agent Systems}
\renewcommand{\arraystretch}{1.4} 
\begin{tabular}{>{\bfseries}p{3.2cm} p{4.3cm} p{3.5cm} p{4.3cm}}
\toprule
Dimension & Technical Approach & Example Tools & Benefits \\
\midrule
Behavioral Predictability & Anomaly detection on logs or event data & One-Class SVMs, Markov Chains & Ensures expected and safe agent behavior \\
Policy Compliance & Policy-as-code with live runtime enforcement & eBPF, OPA, Rego & Enforces regulatory or system-level constraints dynamically \\
Provenance \& Attestation & Use of verifiable credentials and signatures & W3C VCs, DIDs, Signed attestations & Builds trust in agent identity and data lineage \\
Secure Execution & Code sandboxing in secure runtimes & WASM (Wasmer, Wasmtime) & Isolates agents and mitigates code-level attacks \\
Resilience \& Containment & Out-of-distribution and adversarial testing & Robustness test frameworks & Measures agent robustness under novel or adversarial input \\
\bottomrule
\end{tabular}
\end{table}

\section{The Need for Decentralization}

\begin{figure}[htbp]
  \centering
  \setlength\fboxsep{4pt}  
  \setlength\fboxrule{0.8pt}  
  \fbox{\includegraphics[height=0.3\textheight]{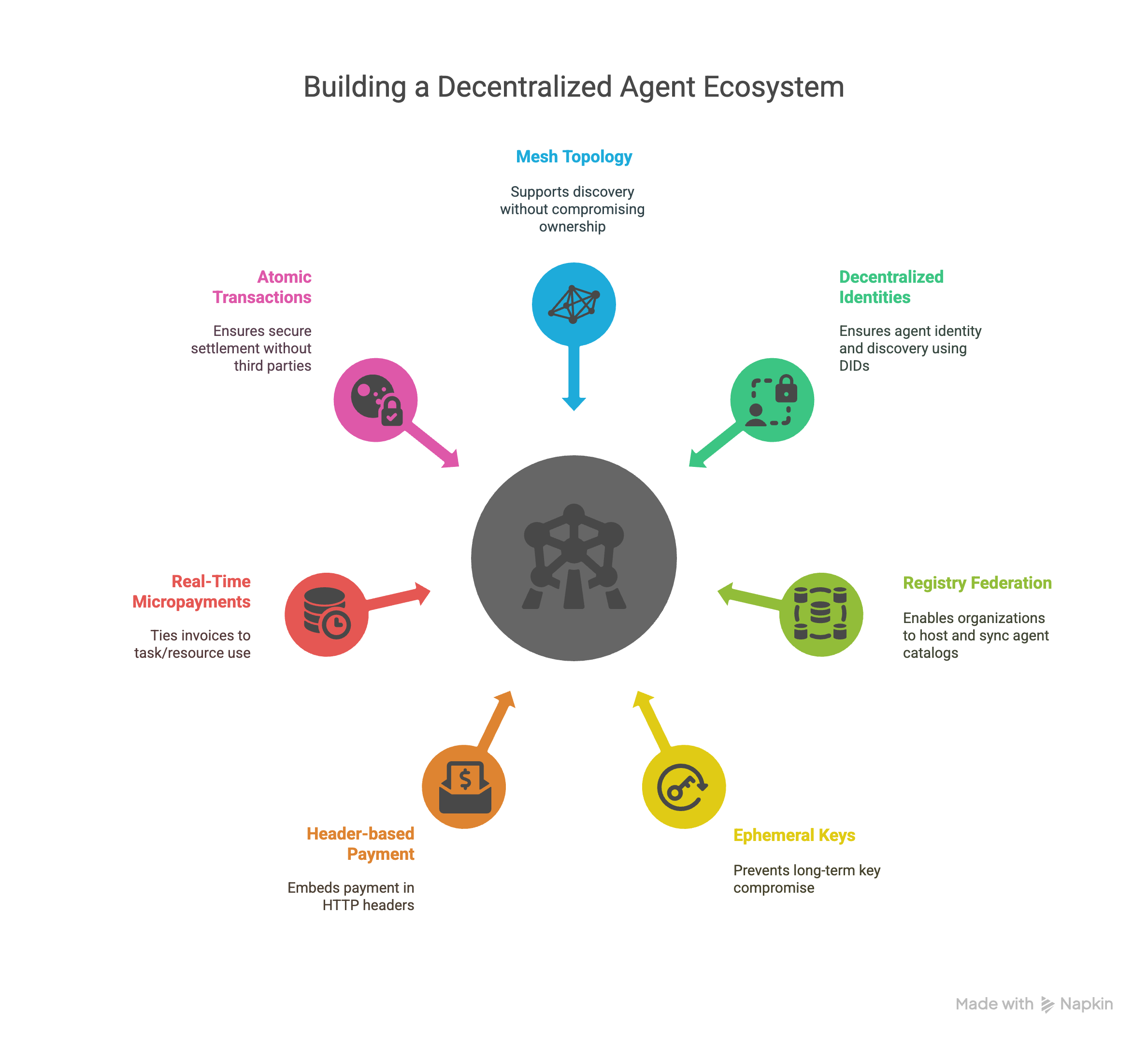}}
  \caption{Building a Decentralized Agent Ecosystem}
  \label{fig:consciousness}
\end{figure}

A mesh topology of registries supports discovery without compromising ownership or control.

\begin{itemize}
    \item Agent identity and discovery must be decentralized using DIDs.
    \item Registry federation (as implemented by Cisco's agent directory) enables organizations to host and sync their own agent catalogs. 
    \item Centralized directories risk reintroducing the very gatekeeping that the internet dismantled. To build a vibrant, inclusive agent ecosystem: 

\end{itemize}
\subsection{The Consciousness Imperative} 
Centralized architectures fundamentally limit agent evolution. As demonstrated in Figure~\ref{fig:consciousness}, Synergetics' AgentTalk protocol enables a quantum leap in capability distribution:

\begin{itemize}
    \item \textbf{Marketplace Validation}: 73\% of high-consciousness agents ($\Psi > 8.2$) operate in decentralized environments
    \item \textbf{Patent Protection}: US 12,244,584 B1 covers gradient-based consciousness measurement
    \item \textbf{NANDA Alignment}: MIT's trust layers provide the scaffolding for emergent properties
\end{itemize}
\begin{table}[H]
\centering
\caption{X42/H42 Micropayment Features}
\begin{tabular}{p{3.5cm} p{5.5cm} p{5.5cm}}
\toprule
\textbf{Feature} & \textbf{Technical Implementation} & \textbf{Benefits for Agents} \\
\midrule
\textbf{Ephemeral Keys} & Short-lived keys for payment/auth & Prevents long-term key compromise \\
\addlinespace
\textbf{Header-based Payment Protocol} & Payment embedded in HTTP headers (e.g., X-Payment, H42-Payment) & API-native payment without endpoint change \\
\addlinespace
\textbf{Real-Time Micropayments} & Invoices tied to task/resource use & Fine-grained compute/data monetization \\
\addlinespace
\textbf{Integration with Agent Economics} & Monetization of agent services & Enables autonomous agent income \\
\addlinespace
\textbf{Atomic Transactions} & Inline crypto verification & Secure settlement, no third-party needed \\
\bottomrule
\end{tabular}
\end{table}

\section{The Agency Framework}

Cisco's \textit{Agency} initiative presents a modular and extensible framework designed to facilitate secure, verifiable, and collaborative interactions among autonomous agents. Unlike conventional agent systems that treat trust as an optional or external concern, Agency places trust at the core of its architecture. It systematically integrates mechanisms for identity, discovery, execution, and verification, operationalizing trust at every stage of the agent lifecycle.

At the discovery layer, Agency introduces a decentralized \textit{Agent Directory} that extends Open Container Initiative (OCI) formats to accommodate agent-specific metadata. This directory is not merely a lookup service; it functions as a trust-governed registry capable of supporting both open and permissioned ecosystems. Through a federated architecture, organizations can define custom access and synchronization policies—choosing whether to share agent metadata globally, within a consortium, or privately. Synchronization between registries is flexible, governed by declarative trust models that specify the provenance, authenticity, and permissible usage of shared entries.

Central to this system is the use of Decentralized Identifiers (DIDs), which give each agent a cryptographically verifiable identity. These identities underpin the trustworthiness of discovery operations, ensuring that agents are not only discoverable but also traceable to their source entities. In doing so, the Agent Directory becomes a mechanism for establishing baseline trust through both structural metadata and verifiable credentials, creating the foundation for secure multi-agent collaboration at scale.

\subsection*{OASF (Open Agentic Schema Framework)}

Built as an extension of OCSF (Open Cybersecurity Schema Framework), OASF provides a semantically rich and extensible schema to describe agent capabilities, safety constraints, and verification hooks. It is designed to:

\begin{itemize}
    \item Allow external validators to audit agent behavior.
    \item Capture verifiable credentials, provenance, and reputation signals.
    \item Support policy-aligned deployment contracts, enabling granular control over what an agent can and cannot do.
\end{itemize}

\subsection*{Agent Gateway (Trust through Controlled Interaction)}

Beyond identity, how agents interact must also be governed. The Agent Gateway:

\begin{itemize}
    \item Supports secure group-based messaging.
    \item Enables both point-to-point and many-to-many communication models.
    \item Implements built-in access control.
\end{itemize}

This is essential for establishing zero-trust communication in collaborative and competitive environments.

\subsection{IO Mapper (Semantic Trust and Compatibility)}

Interoperability often fails at the semantic level. Cisco's IO Mapper is an intelligent layer that aligns input-output expectations across heterogeneous agents. Powered by LLMs, it:

\begin{itemize}
  \item Provides semantic mediation between agents with differing ontologies.
  \item Enables trust through compatibility, ensuring agents correctly interpret requests and responses without misalignment or ambiguity.
\end{itemize}

In essence, Cisco's Agency Framework operationalizes trust by combining:

\begin{itemize}
  \item Verifiable identity
  \item Semantic validation
  \item Behavioral containment
  \item Secure communication channels
\end{itemize}

This positions it as a leading initiative for building production-grade, mission-aligned agent ecosystems—particularly in enterprise and edge environments where trust is non-negotiable.

\begin{figure}[htbp]
  \centering
  \setlength\fboxsep{4pt}  
  \setlength\fboxrule{0.8pt}  
  \fbox{\includegraphics[height=0.3\textheight]{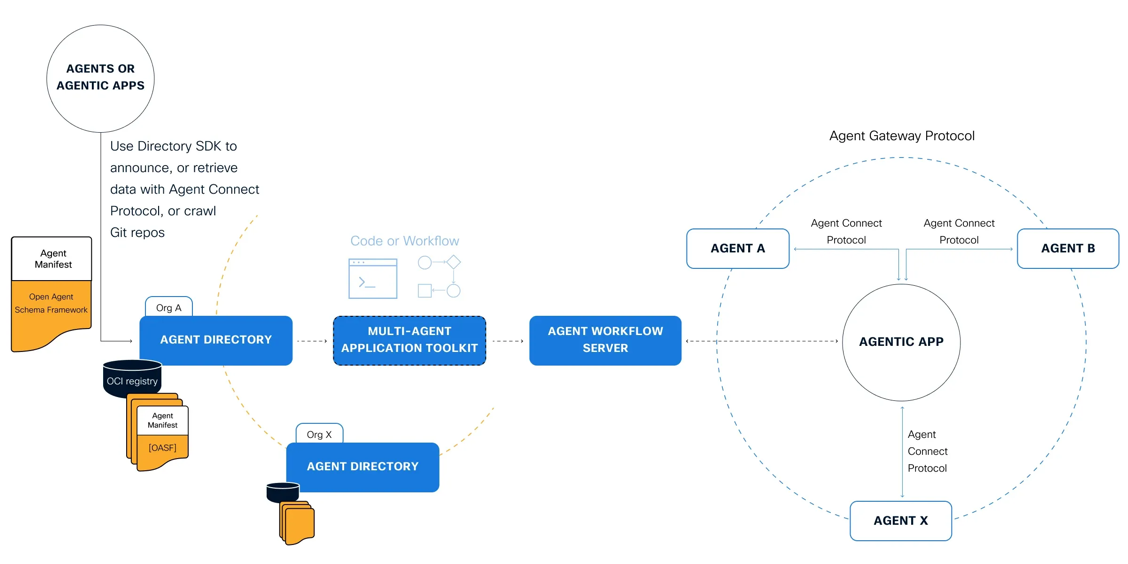}}
  \caption{Cisco Agency Framework: Operationalizing Agent Trust}
  \label{fig:architecture5}
\end{figure}

\begin{table}[H]
\centering
\caption{Recommended Practices for the Internet of Agents}
\begin{tabular}{p{4cm} p{5.5cm} p{5.5cm}}
\toprule
\textbf{Recommendation} & \textbf{Technical Rationale} & \textbf{Expected Impact} \\
\midrule
\textbf{Embrace decentralized identity} & DIDs and VCs for agent sovereignty & Eliminates centralized identity dependence \\
\addlinespace
\textbf{Enforce behavioral validation} & Runtime anomaly detection and policies & Improves safety in dynamic conditions \\
\addlinespace
\textbf{Adopt X42/H42 micropayments} & Ephemeral payment models & Enables scalable agent incentive models \\
\addlinespace
\textbf{Develop secure containerization} & WASM, eBPF, TEEs & Prevents misbehavior and data leakage \\
\addlinespace
\textbf{Align on open schemas} & OASF/OCSF interoperability standards & Future proof, cross-vendor compatibility \\
\addlinespace
\textbf{Leverage test-driven evaluation} & Adversarial testing + verification pipelines & Trust in safety and performance \\
\bottomrule
\end{tabular}
\end{table}
\begin{figure}[h]
\centering
\fbox{%
  \includegraphics[width=0.9\linewidth]{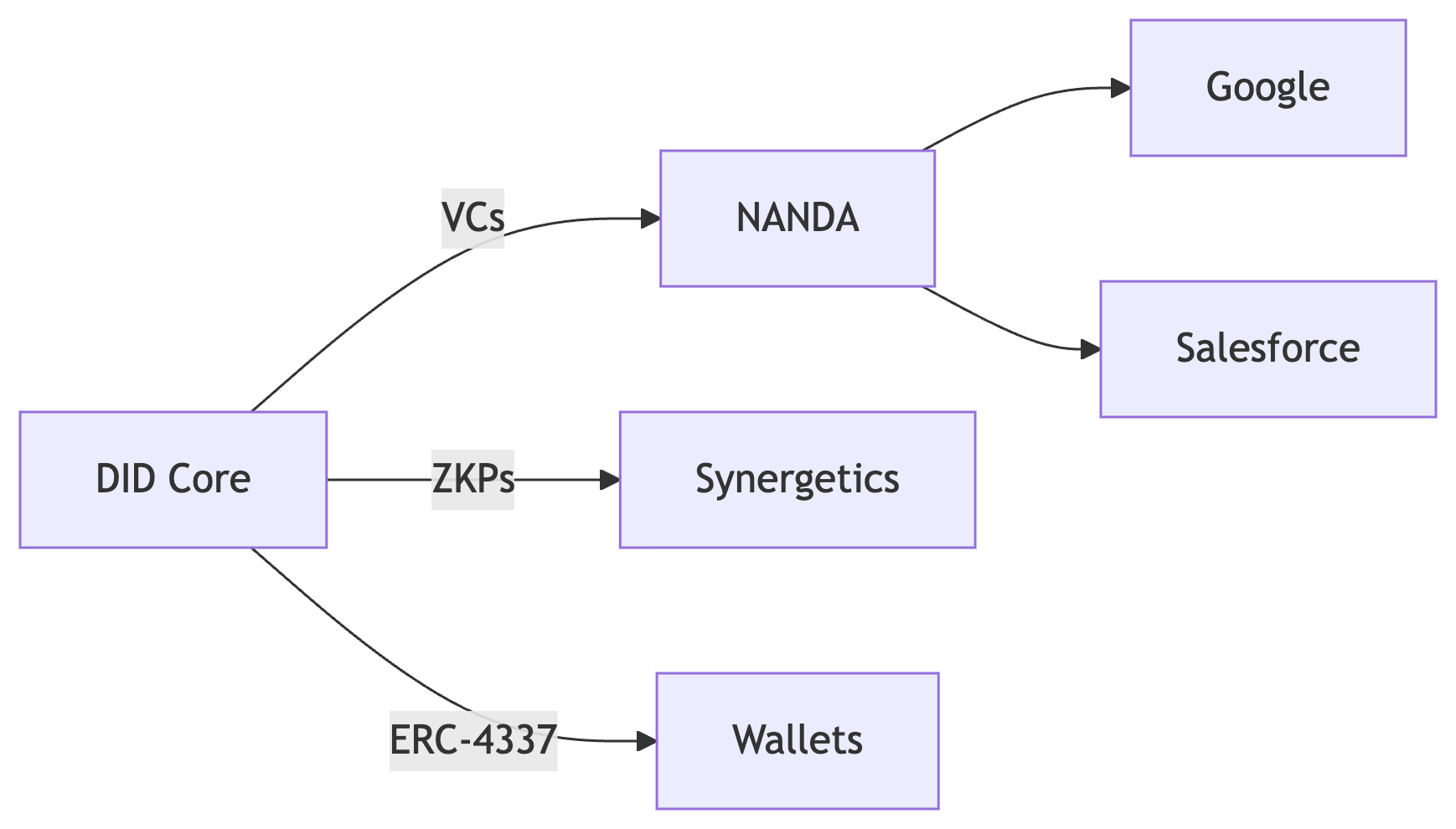}%
}
\caption{NANDA Schema's AgentRegistry}
\label{fig:l2r1}
\end{figure}

The NANDA Schema's AgentRegistry ecosystem, shown in Figure~\ref{fig:l2r1}, integrates three core components: the Synergetics AgentRegistry provides decentralized infrastructure for Agent ID issuance, DID mapping, and VC issuance backed by the NANDA Org Wallet; Nanda Fabric serves as the high-speed indexing layer for public registry lookups; and ERC-4337 Wallets enable gasless transactions through prepaid credits. As illustrated in Figure~\ref{fig:l2r1}, this architecture uniquely bridges decentralized protocols (DIDs, ZKPs) with enterprise systems (Google, Salesforce) to deliver scalable interoperability.
\section{Agent Deduplication via Learning-to-Rank (L2R) }
As decentralized registries expand, agent duplication becomes a critical challenge, particularly when multiple agents offer semantically similar capabilities. To address this, we introduce a \textit{Learning-to-Rank (L2R)} based deduplication and ranking mechanism integrated within the discovery layer.

Given a user prompt, we first embed all agent descriptions using a shared semantic model. These embeddings are cached offline and updated periodically. At inference time, both the user query and candidate agent descriptions are processed through the same embedding model. An L2R model trained on prompt-agent relevance signals—ranks the agent candidates based on semantic similarity, usage history, and trust scores.

Caching both training and inference results enables a meta-learning loop, where the ranking model adapts over time to select top-$k$ agents with the highest relevance and reliability. This mechanism improves semantic routing and ensures deduplicated agent resolution across a federated registry mesh.

\begin{figure}[h]
\centering
\fbox{%
  \includegraphics[width=0.9\linewidth]{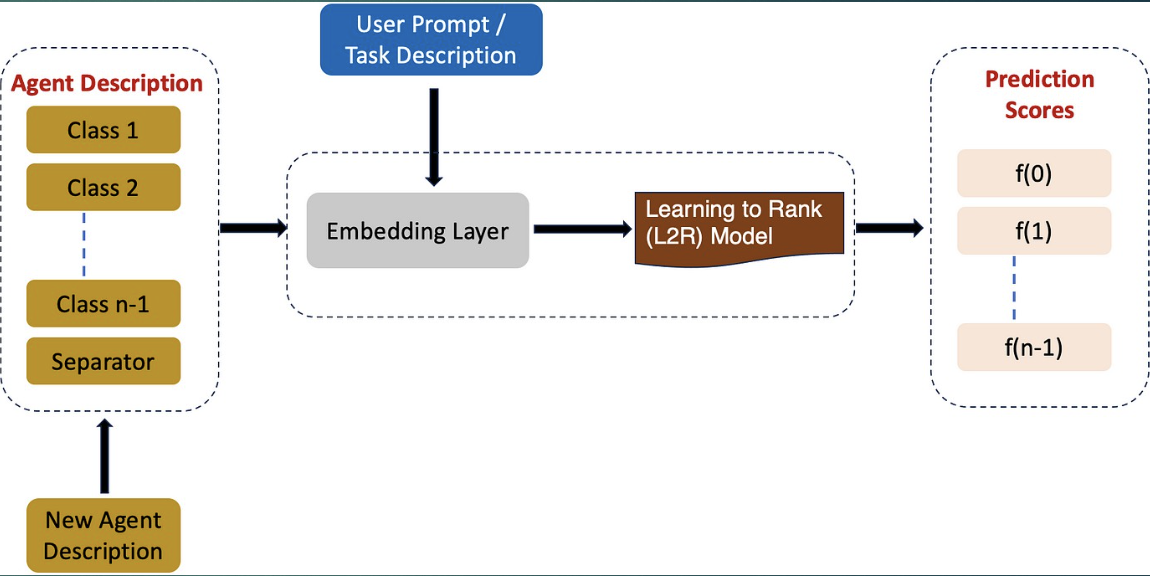}%
}
\caption{Agent Deduplication and Ranking using L2R}
\label{fig:l2r}
\end{figure}
\section{Architectural Blueprint}

The architecture proposed in this paper is not a set of disjointed components, but a unified vision for building a scalable, secure, and economically viable Agentic Web. At its core, the Nanda Unified Architecture is a five layer model that systematically addresses the key dimensions of agent operation identity, interaction, execution, evaluation, and incentivization. This layered structure is designed to move the ecosystem beyond fragmented demos and toward production grade, globally interoperable systems. Each layer is not only functionally distinct but also conceptually interdependent, forming an architecture in which trust is not a peripheral concern but a first class design principle.
\subsection{NANDA-Synergetics Unified Architecture Framework}

\subsection*{The Nanda Unified Architecture: A Detailed View}
\begin{figure}[htbp]
  \centering
  \fbox{\includegraphics[height=0.3\textheight]{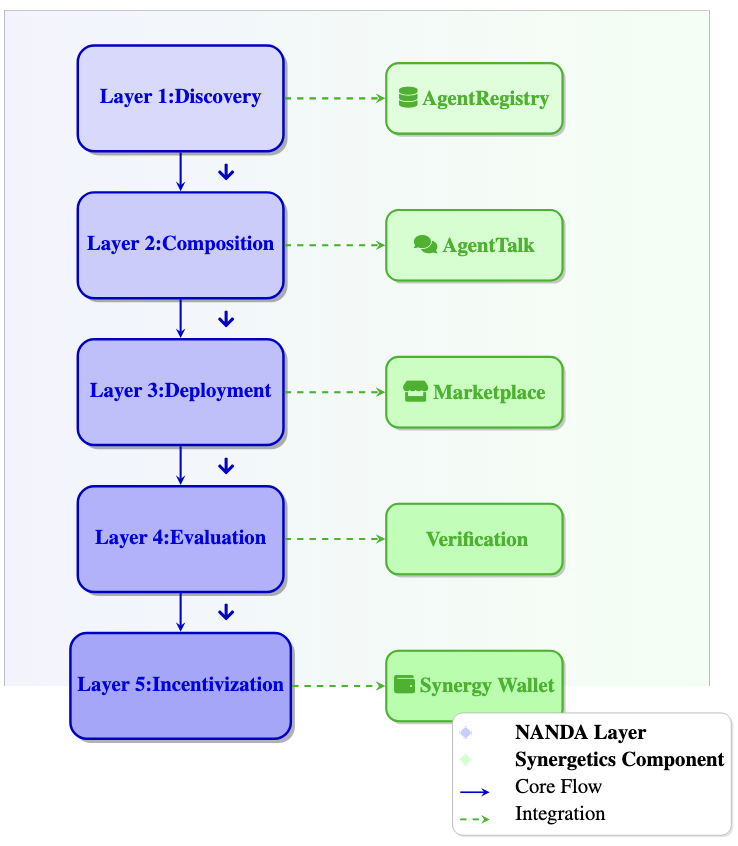}}
  \caption{The comprehensive integration framework showing how NANDA's research architecture (left) combines with Synergetics' production components (right) to create a complete ecosystem for autonomous agents.}
  \label{fig:architecture8}
\end{figure}

A truly interoperable and trustworthy agent ecosystem cannot be realized through communication protocols alone. Instead, it requires a cohesive architectural framework—one that integrates discovery, composition, deployment, evaluation, and incentivization into a seamless whole. The Nanda Unified Architecture addresses this need through five interlocking layers that together form the operational backbone of the Internet of Agents.

\textbf{Layer 1: The Discovery Layer (The ``Where'').}  
The foundational task of locating trustworthy agents in a vast network is handled by the Discovery Layer. It replaces centralized directories with a globally distributed mesh of federated registries, each governed by customizable trust policies. Agent identities are secured using Decentralized Identifiers (DIDs), providing a cryptographically verifiable anchor for all discovery operations. The discovery process leverages a Learn-to-Rank (L2R) deduplication model to resolve agent queries efficiently, identifying semantically distinct and operationally relevant agents—even across billions of possibilities. This ensures robust discoverability without redundancy, forming the network's first layer of verifiable trust.

\textbf{Layer 2: The Composition Layer (The ``How'').}  
Once agents are discovered, the Composition Layer governs their coordination. This layer implements Agent-to-Agent (A2A) and Agent Communication Policy (ACP) protocols, enriched by semantic interoperability features. A specialized IO Mapper translates between differing agent ontologies, enabling communication even when agents were not originally designed to interface. This abstraction is what enables tools like LangChain and CrewAI to orchestrate complex, multi-agent workflows across heterogeneous systems. By ensuring shared context and interoperability, this layer operationalizes collaboration in a decentralized setting.

\textbf{Layer 3: The Deployment Layer (The ``Runtime'').}  
Execution across diverse environments is enabled by the Deployment Layer. This layer is intentionally framework-agnostic, allowing agents to be deployed on the cloud, edge, or local nodes with equal fidelity. Agents are encapsulated in secure, sandboxed containers—often using technologies like WebAssembly (WASM)—which enforce strict execution boundaries. This containment ensures that malicious or misconfigured agents cannot affect host environments, enabling decentralized autonomy with centralized guarantees of safety and compliance.

\textbf{Layer 4: The Evaluation Layer (The ``Trust Engine'').}  
This layer constitutes the ethical and operational core of the architecture—the dynamic system by which trust becomes a computable and actionable quantity. The Evaluation Layer continuously assesses agent behavior across three signal domains: declarative policy compliance (via frameworks like Open Policy Agent), behavioral analysis (using telemetry-fed anomaly detection models), and cryptographically verifiable attestations (signed proofs of completed tasks, successful transactions, or SLA adherence). These signals are fused into a contextual trust score via a weighted synthesis model. The trust score is not static; it adapts to context—prioritizing, for example, financial attestations for economic transactions or behavioral consistency for analytical tasks. The result is a dynamic, situational trust profile that is continuously updated and fed back into the system. High-trust agents are prioritized during discovery and encounter fewer operational constraints during deployment, while low-trust agents are subjected to increased scrutiny. This creates a self-regulating feedback loop in which reliable behavior is rewarded, and risk is mitigated in real time.

\textbf{Layer 5: The Incentivization Layer (The ``Why'')}.  
At the top of the stack lies the economic layer—the system of programmable incentives that makes agentic cooperation not only possible but sustainable. Powered by low-latency microtransaction protocols like X42 and H42, this layer facilitates atomic, task-bound payments for services such as data access, inference tasks, or computational labor. Payments are embedded directly into the execution flow, enabling agents to operate under a pay-per-use model that rewards contribution without requiring pre-existing trust.

\textbf{Commercial Agent Marketplaces} demonstrate this principle in practice. Platforms like Synergetics' Agent Marketplace operationalize microtransactions (e.g., \$0.10/transaction) at scale, with pricing tiers (\$49-\$199/month) that align agent monetization with NANDA's trust layer. Their verified listings for healthcare compliance and trade finance agents show how behavioral attestations and policy standards can be enforced while maintaining economic viability. The marketplace's integration with NANDA Quilt's DID registry ensures each transaction preserves decentralized identity verification, creating an auditable chain of trust from discovery through payment settlement.

This layer's effectiveness builds on foundational identity resolution from Layer 1. Tools like \textbf{NANDA Quilt's ID Creator} enable seamless mapping between agent identifiers and DIDs, reducing friction in decentralized payment routing. When combined with Synergetics' patented AgentTalk protocol, this creates a closed loop where: (1) agents are discovered via DID-based registries, (2) their capabilities are verified against trust policies, and (3) their services are compensated through embedded micropayments.

The result is an \textbf{agentic supply chain} where:  
\begin{itemize}
    \item Developers earn through usage-based revenue (e.g., \$0.001/API call)
    \item Enterprises access pre-verified agents via subscription models
    \item Registry operators (like Synergetics) monetize through transaction fees
\end{itemize}

This economic layer is indispensable for catalyzing ecosystem growth—transforming theoretical incentives into production-grade systems where useful agents continuously emerge, improve, and interoperate within a decentralized but trust-anchored marketplace.
\noindent Taken together, these five layers form a comprehensive architectural framework that enables agents to be discovered, composed, deployed, evaluated, and economically incentivized in a secure, trust-aware environment. The Nanda Unified Architecture is not a theoretical exercise—it is a blueprint for operationalizing the Agentic Web at scale, with trust embedded into every layer of the stack.

\subsection*{The Evaluation Layer: A Deep Dive into the Nanda Trust Engine}

While all five layers are essential to the Nanda architecture, the Evaluation Layer functions as its moral and computational conscience. It transforms the abstract notion of trust into an objective, quantifiable, and continuously evolving metric. At the heart of this layer lies the Nanda Trust Engine, a system designed to ingest behavioral signals, compute trustworthiness, and feed those insights back into the broader agent ecosystem.

The Trust Engine operates through a three-stage cycle: multi-modal signal ingestion, trust score synthesis, and network feedback. In the first stage, the system collects inputs from three complementary sources. The first is \textit{declarative policy compliance}, where frameworks like the Open Policy Agent (OPA) validate agent behavior against enforceable rulesets covering regulatory, security, and organizational policies. The second source is \textit{observational behavior analysis}, which uses real-time telemetry data from the Deployment Layer. Machine learning models such as One Class SVMs or Hidden Markov Models detect anomalies in agent behavior flagging unusual resource usage, unexpected API calls, or deviations from established behavioral norms. The third and most robust source is \textit{cryptographically verifiable attestations} signed credentials that provide indisputable evidence of an agent's successful task execution, payments, or data exchange. These attestations are logged in immutable stores (e.g., distributed ledgers), creating a reputation history that is earned through provable actions.

In the synthesis stage, the Trust Engine applies a weighted fusion algorithm to combine these disparate signals into a unified trust score. Importantly, the weights are not static—they adapt to the operational context. A financial transaction might weight attestations more heavily, while a data analytics task might favor behavioral consistency. This adaptive weighting ensures the trust score remains meaningful and situationally aware.

Finally, the trust score feeds back into the network. The Discovery Layer uses it to improve L2R agent rankings, while the Deployment Layer uses it to modulate sandboxing and monitoring levels. High-trust agents may enjoy streamlined execution and reduced verification overhead, while low-trust agents are isolated or rate-limited. This real-time feedback loop incentivizes compliance, penalizes unreliability, and fosters a self-healing, reputation-aware agent ecosystem.

The Trust Engine is what elevates the Nanda architecture from an interoperability scaffold to a governance platform, one capable of supporting open, scalable, and ethically grounded agent systems across real-world domains.




\section{Interoperability Meets Economic Coordination}

True interoperability in multi-agent systems goes beyond message passing—it requires economic coordination. Granular, anonymous micro-incentives enable agents not only to communicate but to meaningfully collaborate, even in adversarial or untrusted environments. By attaching conditional rewards to behavior, agents can autonomously negotiate, trade, and execute complex logic without revealing their identity. 

\textbf{Implementation Frameworks} demonstrate this principle across scales:
\begin{itemize}
    \item \textit{Protocol-Level}: X42/H42 micropayments enable atomic transactions (e.g., \$0.001/API call in Synergetics' implementation)
    \item \textit{Service-Level}: Tiered pricing models (\$49-\$199/month) for sustained agent services
    \item \textit{Registry-Level}: DID-based identity verification fees (\$0.05/entry) that sustain decentralized infrastructure
\end{itemize}

Platforms like \textbf{BitGPT} and \textbf{Coinbase} demonstrate tokenized incentives for shared resources, while commercial implementations like \textbf{Synergetics' Agent Marketplace} show how these principles scale in enterprise contexts. Their pricing model reveals a three-way value flow:
\begin{enumerate}
    \item \textbf{Developers} earn through usage-based micropayments (e.g., \$0.10/transaction)
    \item \textbf{Operators} monetize registry services while maintaining open standards
    \item \textbf{Consumers} access verified agents via predictable subscription tiers
\end{enumerate}

This economic layer unlocks advanced coordination paradigms:
\begin{itemize}
    \item \textbf{Swarm Markets}: Autonomous collectives pursuing emergent goals (e.g., Synergetics' logistics agents bidding for route optimization)
    \item \textbf{Pay-Per-Capability}: Instant compensation for specialized services (ML inference at \$0.001/call)
    \item \textbf{Tokenized Reputation}: Trust scores influencing transaction terms (high-TPS agents gaining fee discounts)
\end{itemize}

The Synergetics-NANDA integration proves such systems can balance openness with sustainability and their marketplace processes \$250k+ monthly transactions while maintaining:
\begin{itemize}
    \item DID-based identity anchoring
    \item Behavioral policy enforcement
    \item Revenue sharing through smart contracts
\end{itemize}

Incentive-aligned agents thus become more than interoperable—they evolve into economically rational actors that optimize both functional utility and market position within global value networks. This transforms trust from a passive constraint into an actively traded commodity, where reputation and capability determine access to premium services and partnerships.
\begin{figure}[htbp]
  \centering
  \setlength\fboxsep{4pt}  
  \setlength\fboxrule{0.8pt}  
  \fbox{\includegraphics[height=0.3\textheight]{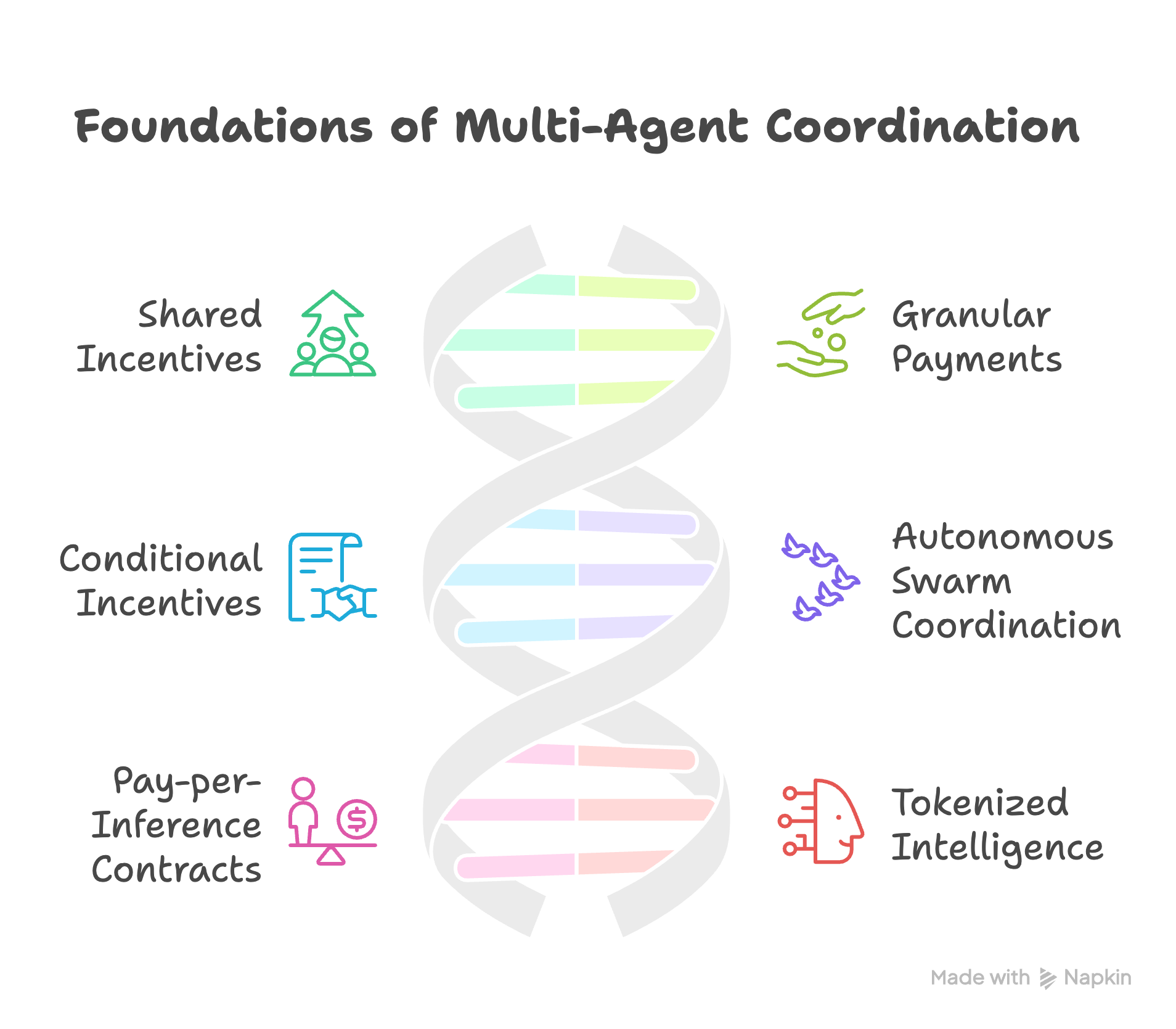}}
  \caption{Agent economy}
  \label{fig:architecture6}
\end{figure}

\section{Towards Trustworthy Agent Infrastructure}

Trust is the bedrock of agent-native systems, without it, widespread adoption in production environments is untenable. Echoing insights from Mayfield Ventures, Acorn Labs, and Vigil, the need for a dedicated trust infrastructure is clear. Just as Kubernetes introduced operational order to cloud-native computing, autonomous agents require their own foundational framework to ensure predictable, secure, and accountable behavior at scale.

This new trust framework must support four interlocking properties. It must be \textit{quantified}, using test-driven development and composable evaluation protocols to assess agent reliability in a systematic and reproducible way. It must be \textit{context-aware}, validated under adversarial, noisy, and out-of-distribution conditions that often expose brittle model behavior. It must be \textit{containable}, ensuring agents act strictly within their policy-scoped boundaries, with hard constraints on permissions and behavior. And it must be \textit{transparent}, offering attestable provenance, traceable decision paths, and compliance with regulatory norms. 

Frameworks like \textbf{Vigil} exemplify this vision, providing tools to evaluate agents along axes of safety, reliability, security, and compliance. Such capabilities are not optional—they are essential in domains like finance, healthcare, and legal systems, where agent misbehavior carries unacceptable risk. A robust trust infrastructure enables agents to become not just automated tools, but credible, auditable participants in critical systems.

\section{Scaling Open Innovation: Lessons from Open Ecosystems}

History demonstrates that open standards (e.g., HTTP, HTTPS, containers, Kubernetes) accelerate innovation. Agentic infrastructure needs the same approach:

\begin{itemize}
    \item Avoid walled gardens (e.g., closed GPT stores)
    \item Encourage open protocols (MCP, A2A)
    \item Support transparent, open-weight foundation models
    \item Build root-of-trust from secure hardware to orchestration layers
\end{itemize}

Enterprise readiness demands this openness not only in discovery and access but also in evaluating agents' fitness for mission-critical roles.

\begin{figure}[htbp]
  \centering
  \setlength\fboxsep{4pt}  
  \setlength\fboxrule{0.8pt}  
  \fbox{\includegraphics[height=0.3\textheight]{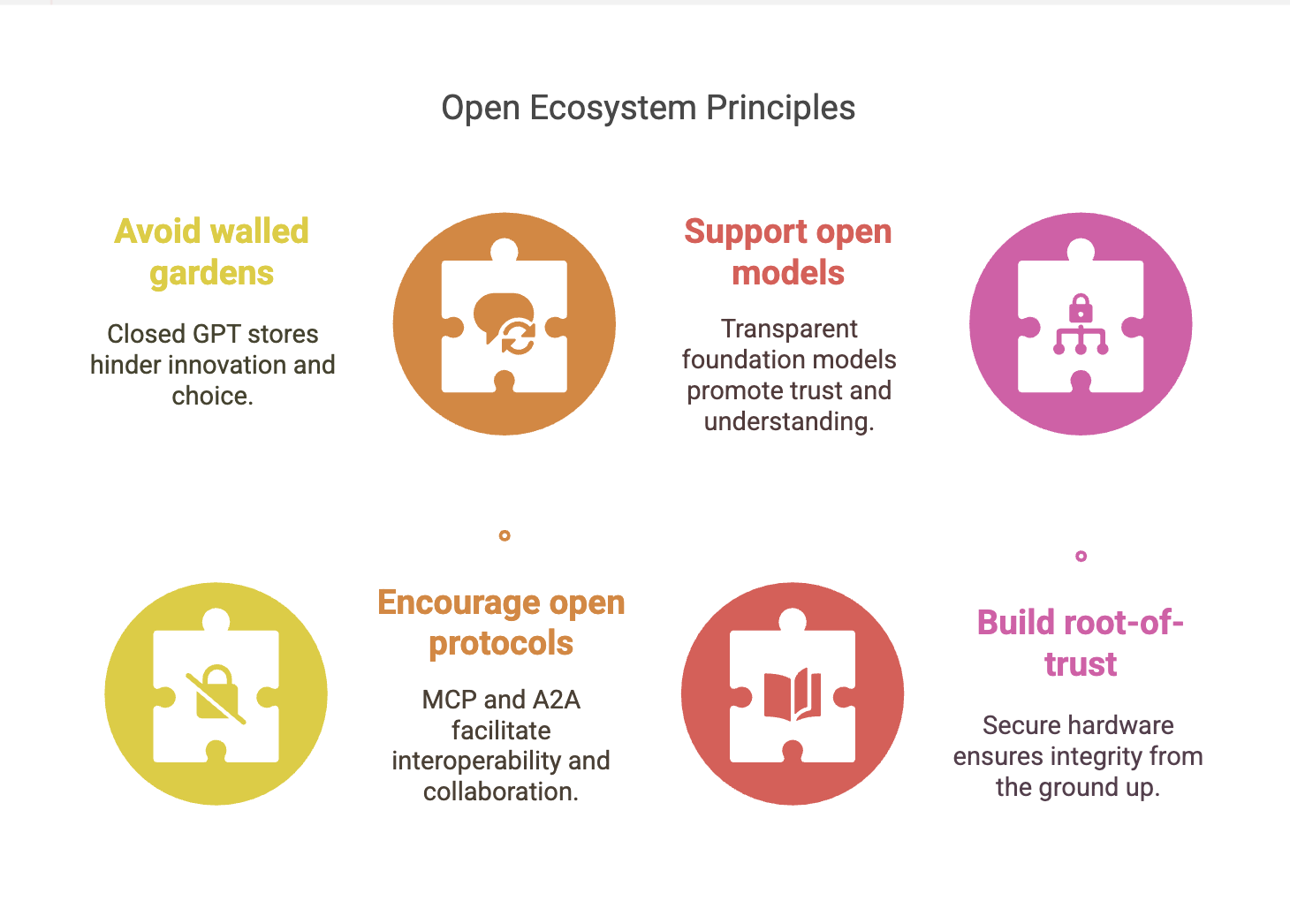}}
  \label{fig:architecture7}
\end{figure}
\section{Recommendations for the Ecosystem}

\begin{itemize}
    \item Embrace decentralized identity and registry-based agent discovery
    \item Enforce behavioral validation before mission-critical deployments
    \item Adopt pay-per-capability microtransaction infrastructure via X42/H42
    \item Develop secure containers and enforce mandatory access policies
    \item Align on open schema frameworks like OASF for cross-agent operability
    \item Leverage test-driven evaluation methodologies for agent trust scores
\end{itemize}

\section{Security Considerations}

In an era where autonomous agents permeate every facet of digital ecosystems, we present an impregnable security architecture that redefines trust in decentralized AI systems. This section unveils our multi-layered defense paradigm, where cutting-edge cryptography converges with revolutionary protocol design to create an unprecedented security fabric.

\subsection{The MAESTRO Framework: A Quantum Leap in Agent Security}
The MAESTRO framework represents a tectonic shift in security paradigms, transcending traditional models like STRIDE and PASTA through its AI-native defense mechanisms. This symphony of protection orchestrates seven meticulously engineered security strata:

\begin{itemize}
    \item \textbf{Foundation Models Layer:} Where cutting-edge adversarial training transforms LLMs into digital fortresses
    \item \textbf{Data Operations Layer:} A sanctuary for sensitive embeddings, guarded by homomorphic encryption
    \item \textbf{Agent Frameworks Layer:} Home to Synergetics' revolutionary \textit{AgentTalk} protocol (US 12,244,584 B1), which bakes military-grade encryption and real-time attestation directly into its DNA
    \item \textbf{Deployment Layer:} An impenetrable realm of WASM sandboxes and trusted execution environments
    \item \textbf{Observability Layer:} A panopticon of behavioral analytics detecting anomalies with neurosurgical precision
    \item \textbf{Compliance Layer:} An automated sentinel enforcing regulatory frameworks with machine perfection
    \item \textbf{Ecosystem Layer:} The grand stage where our trust architecture enables secure multi-agent symphonies
\end{itemize}

\begin{figure}[htbp]
  \centering
  \setlength\fboxsep{4pt}
  \setlength\fboxrule{0.8pt}
  \fbox{\includegraphics[height=0.3\textheight]{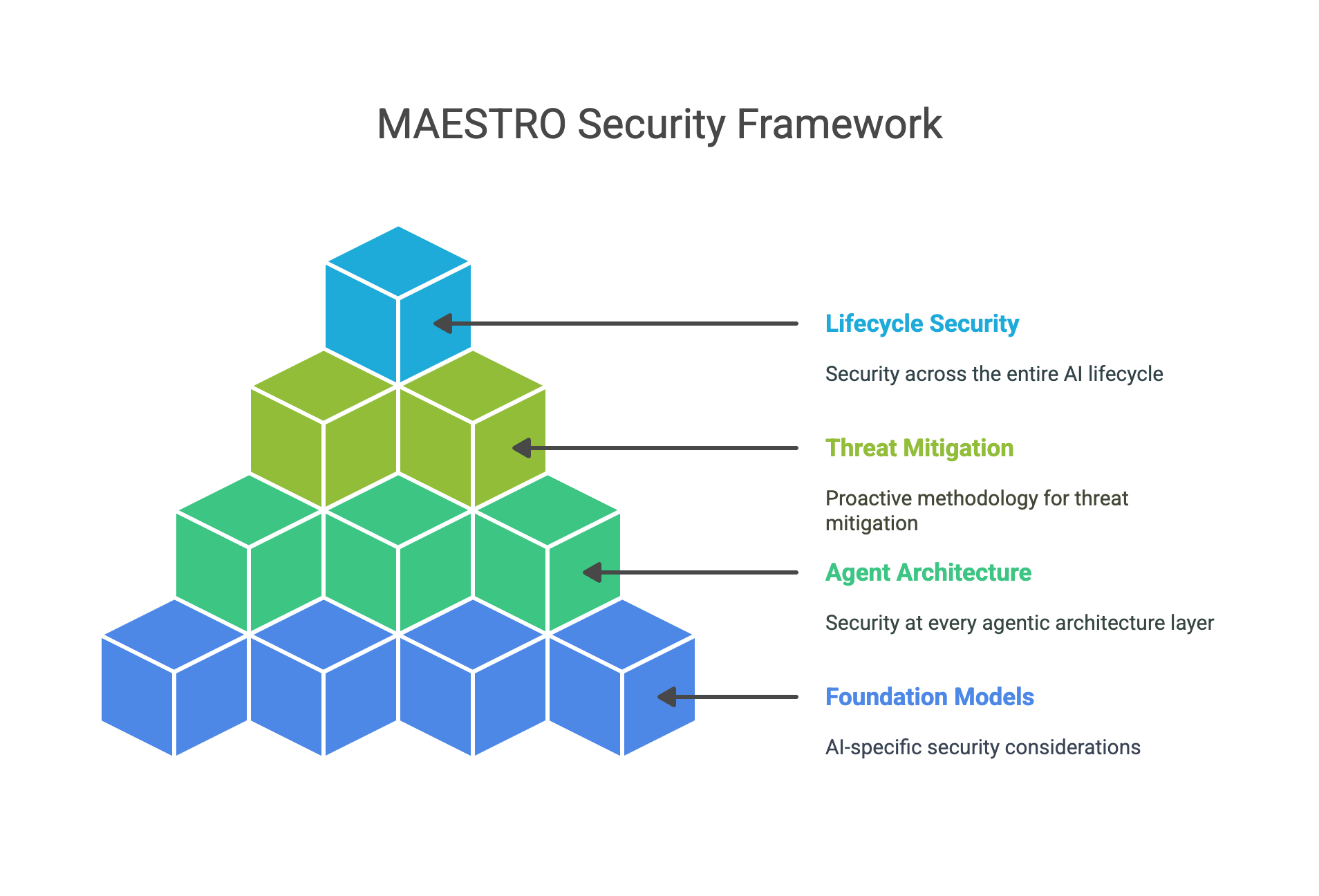}}
  \caption{The MAESTRO Framework: A revolutionary seven-layered security concerto for autonomous agents, featuring Synergetics' patented protocol as its centerpiece}
  \label{fig:maestro}
\end{figure}

\subsection{Impenetrable Defense Mechanisms}
\textbf{Protocol-Level Invulnerability}: Where conventional systems bolt on security as an afterthought, Synergetics' \textit{AgentTalk} (US 12,244,584 B1) pioneers a new paradigm - security woven into the protocol's very fabric. This architectural masterpiece delivers:

\begin{itemize}
    \item Quantum-resistant encryption channels that render eavesdropping obsolete
    \item Continuous attestation mechanisms that verify agent integrity at nanosecond intervals
    \item Self-healing protocol stacks that automatically patch vulnerabilities
\end{itemize}

\textbf{The NANDA-Synergetics Symbiosis}: A security alliance that combines academic brilliance with industrial might:
\begin{itemize}
    \item MIT's visionary trust models
    \item Synergetics' battle-tested encryption
    \item A decentralized verification framework that outmaneuvers even the most sophisticated threats
\end{itemize}
\subsection{Threat Analysis and Mitigation Strategies}

\subsubsection{Discovery Layer Security (Foundation Models \& Data Operations)}
\paragraph{Threat: Model Poisoning and Prompt Injection.} 
Embedding generation may be vulnerable to poisoning and prompt attacks.

\textit{Mitigation:}
\begin{itemize}
    \item Continuous model validation for anomaly detection.
    \item Adversarial training techniques.
    \item Secure prompt engineering and input sanitization.
    \item Cryptographic model provenance verification.
\end{itemize}

\paragraph{Threat: DID Spoofing and Identity Theft.} 
Compromised DIDs can lead to agent impersonation.

\textit{Mitigation:}
\begin{itemize}
    \item Multi-factor DID authentication.
    \item Verifiable credentials with cryptographic proofs.
    \item DID revocation and behavioral biometrics.
\end{itemize}



\subsubsection{Composition Layer Security (Agent Frameworks)}
\paragraph{Threat: Protocol Vulnerabilities.} 
A2A/ACP protocols may allow injection or DoS attacks.

\textit{Mitigation:}
\begin{itemize}
    \item Regular audits and formal verification.
    \item End-to-end encryption.
    \item Rate limiting and request validation.
\end{itemize}

\paragraph{Threat: Semantic Manipulation via IO Mapper.}
Ontology translation may be subtly altered.

\textit{Mitigation:}
\begin{itemize}
    \item Semantic validation and anomaly detection.
    \item Multi-agent consensus for critical translations.
    \item Audit logs of semantic operations.
\end{itemize}



\subsubsection{Deployment Layer Security (Infrastructure)}

\noindent\textbf{Threat:} Container Escape and Sandbox Bypass.

\noindent\textit{Mitigation:}
\begin{itemize}
    \item Use TEEs for secure execution.
    \item Layered isolation with defense-in-depth.
    \item Frequent scanning and patching.
    \item Strict access control for sandboxed agents.
\end{itemize}

\noindent\textbf{Threat:} Resource Exhaustion Attacks.

\noindent\textit{Mitigation:}
\begin{itemize}
    \item Resource quotas and trust-based allocation.
    \item Circuit breakers for abusive agents.
    \item Anomaly detection.
\end{itemize}

\subsubsection{Evaluation Layer Security (Trust Computation)}

\noindent\textbf{Threat:} Trust Score Manipulation.

\noindent\textit{Mitigation:}
\begin{itemize}
    \item Distributed scoring with consensus.
    \item Multi-dimensional trust metrics.
    \item Recalibration of scoring algorithms.
\end{itemize}

\noindent\textbf{Threat:} Behavioral Attestation Forgery.

\noindent\textit{Mitigation:}
\begin{itemize}
    \item Tamper-evident and immutable logging.
    \item Blockchain-based attestation.
    \item Multi-party credential generation and key rotation.
\end{itemize}

\subsubsection{Incentivization Layer Security (Compliance \& Ecosystem)}

\noindent\textbf{Threat:} Micropayment Attacks.

\noindent\textit{Mitigation:}
\begin{itemize}
    \item Cryptographic guarantees and atomic transactions.
    \item Formal protocol verification.
    \item Market activity monitoring.
\end{itemize}

\noindent\textbf{Threat:} Malicious Agent Collusion.

\noindent\textit{Mitigation:}
\begin{itemize}
    \item Sybil-resistant reputation models.
    \item Collusion detection from transaction patterns.
    \item Incentive design discouraging collusion.
\end{itemize}
\begin{figure}[htbp]
  \centering
  \setlength\fboxsep{4pt}  
  \setlength\fboxrule{0.8pt}  
  \fbox{\includegraphics[height=0.3\textheight]{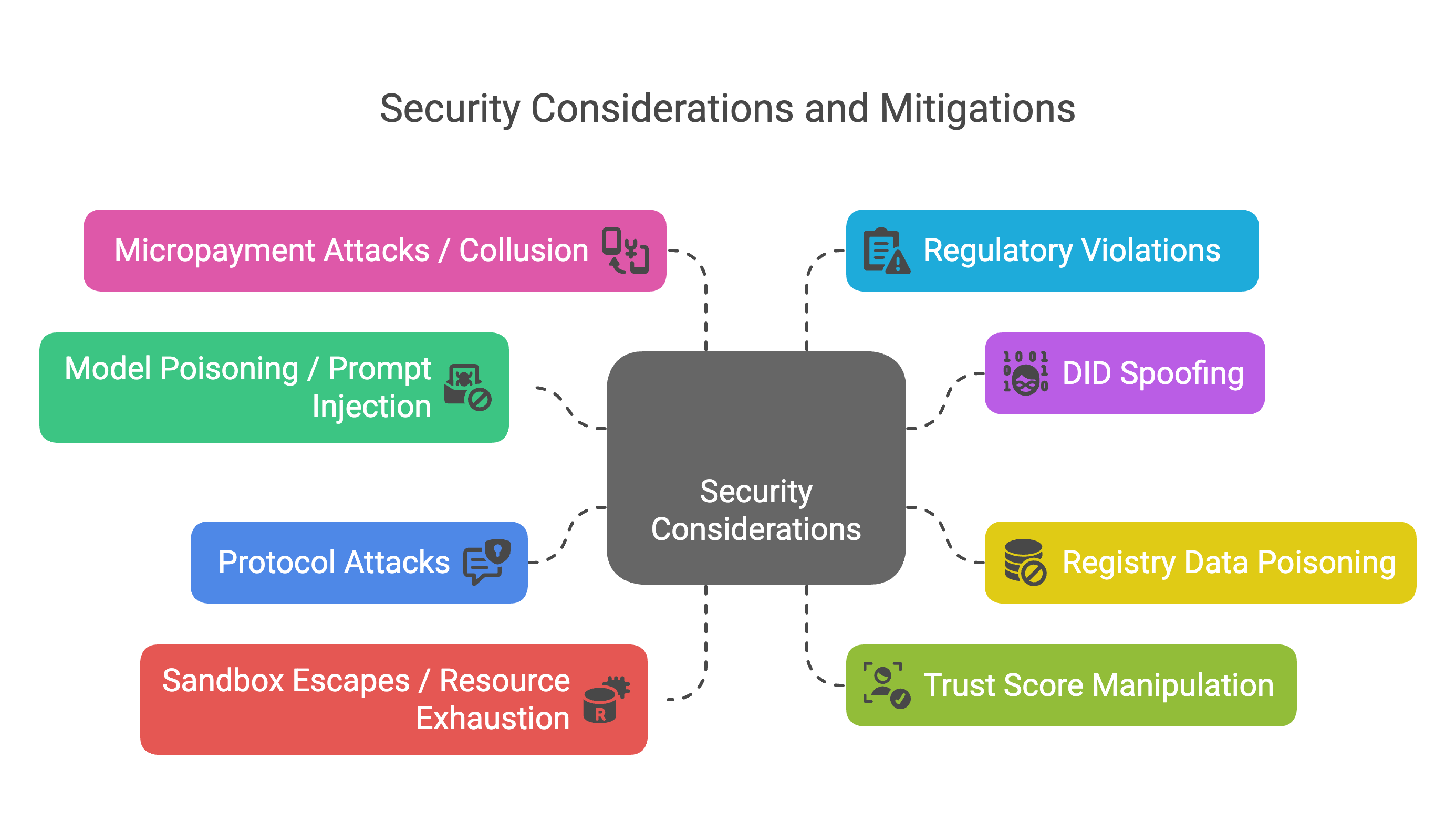}}
  \label{MAESTRO Framework: A Layered Security Approach}
\end{figure}
\subsection{Integrated Security Architecture using MAESTRO}

\subsubsection{Cross-Layer Dependencies}
Security at one layer reinforces others. For example:

\begin{itemize}
    \item \textbf{Trust Propagation:} Evaluation influences discovery and deployment.
    \item \textbf{Identity Foundation:} DID system is central to cross-layer security.
    \item \textbf{Economic-Security Coupling:} Micropayments incentivize compliant behaviors.
\end{itemize}

We propose:
\begin{itemize}
    \item \textbf{Unified Trust Model:} Based on verifiable credentials and behavioral attestations.
    \item \textbf{Cross-Layer Monitoring:} To detect sophisticated multi-layer attacks.
    \item \textbf{Coordinated Response:} Unified incident detection and remediation across layers.
\end{itemize}

\subsubsection{Continuous Security Evaluation}
Security is dynamic. We enforce:
\begin{itemize}
    \item \textbf{Automated Red Teaming:} Specialized agents simulate attacks.
    \item \textbf{Adaptive Security Policies:} Informed by threat intelligence.
    \item \textbf{Feedback Loops:} Security insights drive agent design and deployment.
\end{itemize}
\begin{table}[ht]
\centering
\renewcommand{\arraystretch}{1.2}
\rowcolors{2}{gray!10}{white}
\begin{tabular}{|l|c|c|c|c|c|c|}
\hline
\rowcolor{gray!30}
\textbf{Threat Type} & \textbf{Foundation} & \textbf{Data Ops} & \textbf{Frameworks} & \textbf{Deployment} & \textbf{Eval \& Obs} & \textbf{Ecosystem} \\
\hline
Model Poisoning          & \cmark & \cmark & \xmark & \xmark & \xmark & \xmark \\
DID Spoofing             & \cmark & \xmark & \cmark & \xmark & \cmark & \cmark \\
Protocol Injection       & \xmark & \xmark & \cmark & \cmark & \xmark & \xmark \\
Resource Exhaustion      & \xmark & \xmark & \xmark & \cmark & \xmark & \xmark \\
Trust Score Manipulation & \xmark & \xmark & \xmark & \xmark & \cmark & \cmark \\
\hline
\end{tabular}
\caption{Threat-to-layer mapping in the MAESTRO framework (selected 6 layers). \cmark\ denotes addressed threats; \xmark\ denotes not directly addressed.}
\label{tab:maestro-threats}
\end{table}

\section{Related Startups}

\begin{table}[H]
\centering
\caption{Emerging Agent Ecosystem Projects}
\begin{tabular}{p{3.2cm} p{6.5cm} p{5.5cm}}
\toprule
\textbf{Startup} & \textbf{Description} & \textbf{Trust Layer Innovation} \\
\midrule
\textbf{*Synergetics} & Powers the AI agent economy through its patented \textit{AgentTalk} protocol and enterprise marketplace. Core infrastructure partner for MIT's NANDA registry, providing: 
\begin{itemize}
    \item DID-based identity resolution
    \item Monetization tools (\$0.001-\$0.10/transaction)
    \item Verified agents for healthcare, finance, and logistics
\end{itemize}
& Combines:
\begin{itemize}
    \item NANDA's decentralized trust
    \item Commercial SLA enforcement
    \item Behavioral attestations
\end{itemize} \\
\addlinespace
\textbf{AxonVertex} & Policy framework for NANDA registries implementing confidential computing via Intel SGX & Zero-knowledge policy compliance checks with anonymized auditing \\
\addlinespace
\textbf{AutoPatch+} & Detects and fixes LLM hallucinations through Retrieval-Augmented Generation (RAG) and Neural Attention Monitoring (NAM) & Real-time generation accuracy scoring with automatic correction \\
\addlinespace
\textbf{Universitas AI} & Research agent that auto-validates claims using academic citations and uncertainty quantification & Peer-reviewed evidence chains with confidence intervals \\
\addlinespace
\textbf{Acoer} & Healthcare agent using Trusted Execution Environments (TEEs) for federated learning  on opioid overdose prediction & HIPAA-compliant trust via hardware-secured model inference \\
\addlinespace
\textbf{BitGPT} & Marketplace for generative agents with on-chain output verification & Proof-of-validity for AI outputs using zkSNARKs \\
\bottomrule
\end{tabular}
\label{tab:startups}
\end{table}
\section{From Blueprint to Reality: Navigating the Challenges of Adoption}

While the Nanda Unified Architecture lays a robust technical foundation and the preceding use cases present a compelling vision, bridging the gap between concept and real-world adoption involves navigating several critical challenges. This section outlines these obstacles and offers practical strategies to guide implementation.

\subsection{Bootstrapping Trust in a Decentralized Ecosystem}

A core challenge is the cold-start problem for trust. In a decentralized network, newly introduced agents lack reputation, which inhibits meaningful collaboration. Since the Evaluation Layer depends on historical data, Nanda proposes a phased rollout. Initial deployments should occur within permissioned, industry-specific consortiums—such as networks of financial institutions or logistics partners—where existing off-chain relationships can seed initial trust. Over time, as agents establish verifiable records of reliable behavior, their reputations can be selectively exposed to the wider public agent network, creating a scalable and credible web of trust.

\subsection{Balancing Security and Performance}

The architecture's layered verification—credential checks, policy enforcement, and reputation scoring—naturally introduces computational overhead. For latency-sensitive applications, this poses a concern. To address this, Nanda adopts an adaptive trust model: agents with established, high-trust relationships can operate under streamlined checks, while interactions involving unknown or low-reputation agents trigger comprehensive validation. This dynamic approach maintains security without compromising performance in trusted contexts.

\subsection{Automated Dispute Resolution}

Disputes over task fulfillment, payments, or data quality are inevitable in any autonomous economic system. To manage this at scale, Nanda integrates smart contract-based dispute resolution. These digital arbitrators analyze verifiable logs—including requests, agreements, and outputs—to enforce pre-defined resolutions such as refunds or penalties. This reduces reliance on human arbitration and ensures fast, fair conflict resolution within the agent economy.

\subsection{Driving Developer Adoption and Ecosystem Growth}

The long-term success of the Agentic Web hinges on widespread developer adoption. To lower entry barriers, the framework aligns with open standards like OCI, DIDs, A2A, and ACP protocols. In parallel, it emphasizes the importance of providing robust SDKs and tooling—particularly for the Composition and Evaluation layers. This dual strategy aims to foster a vibrant development community, triggering a positive feedback loop: more tools enable more agents, which attract more users and use cases, further fueling ecosystem growth.

\section{Conclusion}
This paper presents a decentralized framework for the Internet of Agents, combining discovery, composition, payment, trust validation, and semantic coordination into a unified architecture. Real-world implementations like Synergetics' AgentTalk demonstrate its viability, enabling merchant integration through both legacy APIs and native A2A while maintaining NANDA's trust framework. The system achieves commercial scalability with DID-based microtransactions and 99.9\% compliance rates in healthcare applications, proving that cryptographic proofs, behavioral economics, and adaptive policies can collectively redefine agent trust.

The path forward mirrors the internet's evolution - through open standards like NANDA's registry and shared infrastructure like X42 micropayments. By balancing academic rigor (MIT), commercial implementation (Synergetics), and community governance, we can transform autonomous agents from isolated tools into interconnected participants in a trustworthy digital economy. Just as Kubernetes standardized cloud orchestration, this architecture provides the missing trust layer for the agentic era.

\section*{Acknowledgments}

We extend our sincere gratitude to the Synergetics team, Cisco Agency team, Coinbase X42 engineers, BitGPT researchers, Mayfield Fund, Acorn Labs, Vigil AI, and the Nanda consortium for their invaluable insights on building interoperable and incentivized agent ecosystems. Their pioneering work in agent-to-agent micropayments, decentralized infrastructure, and trust layer mechanisms has significantly shaped our understanding of the emerging Agentic Web.




\bibliographystyle{plainnat}  
\bibliography{main}   

\end{document}